\documentclass{article}

\usepackage{arxiv}

\usepackage[utf8]{inputenc} 
\usepackage[T1]{fontenc}    
\usepackage{hyperref}       
\usepackage{url}            
\usepackage{booktabs}       
\usepackage{amsfonts}       
\usepackage{nicefrac}       
\usepackage{microtype}      
\usepackage{graphicx}
\usepackage{doi}
\usepackage{amsmath}  
\usepackage{mathtools}
\usepackage[ruled,vlined]{algorithm2e}
\usepackage{array}
\usepackage{geometry}
\usepackage{multirow}
\usepackage{subcaption}
\usepackage{qcircuit}

\usepackage[square,sort,comma,numbers]{natbib}

\title{Solving Maxwell’s Equations using Variational Quantum Imaginary Time Evolution}

\author{ Nam Nguyen \\
	Integrated Vehicle Systems, Applied Mathematics\\
	Boeing Research \& Technology\\
	Huntington Beach, CA, 92647, USA \\
	\And
	Richard Thompson \\
	Integrated Vehicle Systems, Applied Mathematics\\
	Boeing Research \& Technology\\
	Huntsville, AL, 35824, USA \\
}

\date{}

\hypersetup{
pdftitle={A template for the arxiv style},
pdfsubject={q-bio.NC, q-bio.QM},
pdfauthor={Nam Nguyen, Richard Thompso},
}

\begin{document}
\maketitle

\begin{abstract}
Maxwell's equations are fundamental to our understanding of electromagnetic fields, but their solution can be computationally demanding, even for high-performance computing clusters. Quantum computers offer a promising alternative for solving these equations, as they can simulate larger and more complex systems more efficiently both in time and resources. In this paper we investigate the potential of using the variational quantum imaginary time evolution (VarQITE) algorithm on near-term quantum hardware to solve for the Maxwell’s equations. Our objective is to analyze the trade-off between the accuracy of the simulated fields and the depth of the quantum circuit required to implement the VarQITE algorithm. We demonstrate that VarQITE can efficiently approximate the solution of these equations with high accuracy, and show that its performance can be enhanced by optimizing the quantum circuit depth. Our findings suggest that VarQITE on near-term quantum devices could provide a powerful tool for solving PDEs in electromagnetics and other fields.
\end{abstract}

\section{Introduction}
\label{sec: Introduction}
High-performance computing (HPC) has transformed product design methodology through simulation and modeling of various complex partial differential equations (PDEs) \cite{Tinoco1998_BoeingCFD, Johnson2003ThirtyYO, Tinoco2007_BoeingCFD, Wang2015TowardsHA}. However, the complexity and size of many PDE problems pose significant challenges, making them intractable even for the most powerful HPC systems \cite{jameson2006DNS}. Quantum computing, on the other hand, has demonstrated remarkable efficiency in handling high-dimensional problems, offering exponential speed-up for tasks such as electronic structure calculations of molecular systems \cite{Reiher2017, Bauer2020, santagati2023DrugDesign}, large linear system solving \cite{HHL_2008, CKR, Gilyn2018QuantumSV, costa2022optimalLinearSystem, an2022optimalLinearSystem}, and prime factorization \cite{Shor91}. This has led to growing interest in integrating quantum computing to solve complex PDEs that arise in real-world applications \cite{Lapworth1, Lapworth2, Meng:2023zek}, since most PDE-solvers rely on powerful linear algebra solvers. For instance, if we consider the Poisson equation with Dirichlet boundary conditions: 
\begin{equation}
\label{eq:Poisson}
    -\Delta u(\mathbf{x} ) = f(\mathbf{x} )
\end{equation}
for some $\mathbf{x} \in \Omega$, with $u(\mathbf{x}) =0$ for $\mathbf{x} \in \partial \Omega$ where $\Omega \in \mathbb{R}^d$ with Lipschitz boundary  $\partial \Omega$, and $u$ is a real-valued function $u: \Omega \to \mathbb{R}$ and $f: \Omega \to \mathbb{R}$. In the 1D case, one can use central difference approximation to discretize the second derivative to rewrite it as $\frac{-u_{i-1} + 2u_i - u_{i-1} }{h^2} = b_i$ for $i=1,2, \cdots, N_x -1$ with $u_0 = u_{N_x} = 0$ and $N_x +1$ represents the number of of discretization points with the mesh size equals to $1/N_x$. Also note that $b_i = f(x_i)$ . The discretized formulation can be expressed as a linear system of equations, $A \mathbf{x} = \mathbf{b}$, where $A$ is a tridiagonal matrix of the form
\begin{equation}
A = \frac{1}{h^2} 
\begin{bmatrix}
    2 & -1 & \cdots & 0 &  0 \\
   -1 &  2 & \cdots & 0 &  0 \\
   \vdots & \vdots & \ddots & \vdots & \vdots \\
   0 & 0 & \cdots & 2 & -1 \\
   0 & 0 & \cdots & -1 & 2 \\
\end{bmatrix}
\end{equation}
and $\mathbf{x} = (u_1, u_2, \cdots, u_{N_x - 1})^T$ and $\mathbf{b} =(f(x_1), f(x_2), \cdots , f(x_{N_x -1}) )^T$. The quantum solution to the time-independent linear partial differential equation (PDE) presented in Equation \ref{eq:Poisson} has been extensively studied \cite{Cao_2013, Liu_Poisson, Wang2020}. In the case of time-dependent linear PDEs, particularly when considering PDEs of the form:

\begin{equation}
\label{eq:linear_time_dependent_PDE}
\frac{\partial u}{\partial t} = F(u; \{\partial_i, a_i\}_{i=1:d}) + b(x,t) \quad \forall \mathbf{x} \in \Omega \subseteq \mathbb{R}^d, \quad u(\mathbf{x}, t_0) = u_0
\end{equation}

where $F(u; \{\partial_i, a_i\})$ is a linear combination of spatial differential operators $\{\partial_i\}_{i=1:d}$ and their corresponding parameters $\{a_i\}_{i=1:d}$, and $b(x,t)$ is the source function. An example of Equation \ref{eq:linear_time_dependent_PDE} is the heat equation:

\[
\frac{\partial u}{\partial t} = \alpha \left( \frac{\partial^2 u}{\partial x_1^2} + \frac{\partial^2 u}{\partial x_2^2} + \cdots + \frac{\partial^2 u}{\partial x_d^2} \right)
\]

where $a_i = \alpha$ for all $i$ and is constant through time $t$, with $\alpha > 0$. Any PDEs in the form of Equation \ref{eq:linear_time_dependent_PDE} can also be reformulated as a linear system of equations $\mathcal{A}x = b$, by first performing a semi-discretization (Method of Lines (MOL) \cite{MOL}) to obtain a system of linear ordinary differential equations (ODEs), followed by time-marching in the Feynman's clock representation as proposed in \cite{Berry_2014}. More explicitly, after performing semi-discretization on Equation \ref{eq:linear_time_dependent_PDE}, one obtain the following system of linear ODEs:

\begin{equation}
\label{LinearODE}
\dfrac{d \mathbf{u}}{dt} = A(t) \mathbf{u}(t) + \mathbf{b}(t) \hspace{1.5 cm} \mathbf{u}(t_0) = \mathbf{u}(0)
\end{equation}

where $\mathbf{u}$ and $\mathbf{b}$ are vectors with $N_x + 1$ components, and $A$ is an $(N_x + 1) \times (N_x + 1)$ matrix representing the discretized differential operators that also incorporates the boundary conditions. The matrix $A$ is typically sparse. Note that the MOL is a general method to solve PDEs in the form of Equation \ref{eq:linear_time_dependent_PDE}, for both classical and quantum computing settings \cite{Satofuka1987, Oymak1996, Gaitan2020, Oz2021, Liu2021}. It converts a (system of) PDE(s) into a system of linear ODEs, where well-known stable ODE integrator can be used to the time integration (time marching). The simplest time integrator is the Forward-Euler (an explicit scheme), which updates the solution at time step $t_{n+1}$ ($u(t_{n+1}) = u^{n+1}$) as follows:
\begin{equation}
    \label{eq: Forward Euler}
    u^{n+1} = (I + \Delta t A_n)u^n + \Delta t b_n
\end{equation}
Note that equation \ref{eq: Forward Euler} can be encoded as a linear system of equations $\mathcal{A} \mathbf{x} = \mathbf{b}$ as follow: 
\begin{equation}
\label{eq:FeynmanClock-ForwardEuler}
\overbrace{
\begin{bmatrix}
    I & 0 & 0 & \cdots  &  0 \\
   -(I+ \Delta t A_1) &  I  & 0 &  \cdots & 0 \\
   0 & -(I+ \Delta t A_2)  & I & \cdots & 0 \\
   \vdots & \vdots &   & \ddots & \vdots \\
   0 & 0 & \cdots & -(I+ \Delta t A_{k-1}) & I \\
\end{bmatrix}}^{\mathcal{A}}
\overbrace{
\begin{bmatrix}
    \mathbf{u}^0 \\
    \mathbf{u}^1  \\
    \mathbf{u}^2  \\
   \vdots \\
   \mathbf{u}^k  \\
\end{bmatrix}}^{ \mathbf{x} }
=
\overbrace{
\begin{bmatrix}
    \mathbf{u}(0)  \\
    \Delta t \mathbf{b}_1 \\
    \Delta t \mathbf{b}_2 \\
   \vdots \\
   \Delta t \mathbf{b}_{k-1}  \\
\end{bmatrix}}^{ \mathbf{ \mathcal{ B } }  }
\end{equation} 
where each entry of $\mathcal{A}$ is a block of the dimension of each $A_j = A(t_j)$, and each entry of $\mathbf{x}$ and $\mathbf{ \mathcal{ B } } $ is a block of dimension of each $\mathbf{u}^j = \mathbf{u}(t_j)$. Note that higher order time integrator (both explicit and implicit) schemes can also be reformulated as a linear system of equations $\mathcal{A} \mathbf{x} = \mathbf{b}$. Furthermore, other classical numerical methods like Crank-Nicolson, Lax-Friechrichs, etc. can also be reformulated this way.

For time-dependent non-linear PDEs, in particular, those of the form:
\begin{equation}
\label{eq:Nonlinear_time_dependent_PDE}
\frac{\partial u}{\partial t} = N(u; \{\partial_i, a_i\}_{i=1:d}) + F(u; \{\partial_i, a_i\}_{i=1:d}) + b(x,t) \quad \forall \mathbf{x} \in \Omega \subseteq \mathbb{R}^d, \quad u(\mathbf{x}, t_0) = u_0
\end{equation}
where $F$ is defined as before (a linear operator) and $N$ is a nonlinear operators. Examples of Equation \ref{eq:Nonlinear_time_dependent_PDE} are Viscous Burger's and compressible Navier-Stokes equations. If we perform MOL on Equation \ref{eq:Nonlinear_time_dependent_PDE}, we will obtain a coupled system of nonlinear ODEs, where a time-integrator can then be applied.To reformulate the time integration step into a linear system of equations $\mathcal{A} \mathbf{x} = \mathbf{b}$, as shown previously, we need to utilize a linearization technique from nonlinear dynamical systems, such as Carleman linearization \cite{CarlemanEmbedding} or Koopman operator theory \cite{KoopmanTheory}, to convert the system of nonlinear ODEs into a system of linear ODEs. This system of linear ODEs can potentially be infinite-dimensional, necessitating truncation for practical implementation. This approach is taken by \cite{Liu2021, liu2023NonlinearReactionDiffusion, krovi2023ImproveLinearNonlinearODES, An2023FractionalRxnDiffusion}.

Despite the fact that reformulating the time-marching step as a linear system of equations leads to very large linear systems to be solved, efficient quantum algorithms with exponential speedup have been developed for solving linear systems of equations, making this approach becomes an attractive option for solving PDEs on quantum computers. In particular, the quantum algorithm given by Harrow, Hassidim, and Lloyd (HHL), can solve this problem with runtime in log of the dimension of $\mathcal{A}$ and polynomial in the condition number of $\mathcal{A}$. More concretely, the runtime for HHL is $O(\log(N) s^2 \kappa^2/\epsilon)$ where $N$ is the dimension of $\mathcal{A}$, $s$ is the sparsity of $\mathcal{A}$, $\kappa$ is the condition number of $\mathcal{A}$, and $\epsilon$ is the allowable error. In contrast, Conjugate Gradient (CG), a leading approach for solving sparse linear systems of equations, has a runtime of $O(N s \sqrt{\kappa} \log(1/\epsilon))$ \cite{Conjugate_Gradient}. Many improved quantum algorithms have been proposed to improve the scaling of HHL by, for instance \cite{CKR} used the method of Linear Combination of Unitaries (LCU) to implement the Fourier or Chebyschev fitting of the function $f(x) = 1/x$ to push the runtime to $O(\log(N)s^2 \kappa^2 poly\log(1/\epsilon) )$, and \cite{Gilyn2018QuantumSV} uses quantum singular value transformation (QSVT) to perform the polynomial transformation of the singular values of the linear operator, $\mathcal{A}$, embedded in a larger unitary matrix to get the run time scaling to $O(\log(N) s^2 \kappa^2 \log(1/\epsilon) )$. Recently, further improvement has been achieved through the use of the adiabatic theorem, reducing the condition number scaling from $\kappa^2$ to $\kappa$ \cite{an2022optimalLinearSystem, costa2022optimalLinearSystem}, thus achieving optimal scaling.

Another alternative approach for performing time-marching, without the need to recast it into a linear system of equations, is based on the work of \cite{Fang2023timemarchingbased}, where the exponential integrator (integrating factor) is directly implemented. More explicitly, if we suppose $A(t)=A$ and $\mathbf{b} = 0$, then the solution to \ref{LinearODE} takes the form 
\begin{equation}
    \mathbf{u}(t) = e^{At}\mathbf{u}(0)
\end{equation}
Note that if $A = -iH$ ( anti-Hermitian) then this problem reduces to a quantum simulation. However, in general, $A$ may not even be diagonalizable, and hence QSVT can't be used directly to implement $e^{At}$. A potential remedy to this problem is to realize that $e^{A}$ has a contour integral representation of the form \cite{TrefethenExpOperator}
\begin{equation}
\label{eq: contour_int_eAt}
    e^{A} = \dfrac{1}{2\pi i} \oint_{\Gamma} e^{z}(zI_n-A)^{-1} dz
\end{equation}
where $\Gamma$ is a contour that encompasses all the eigenvalues of $A$ within its interior, and $I_n$ as the identity matrix with same dimension as $A$. Note that one can opt for $\Gamma$ to be a unit circle with a radius of $\beta$, provided $\beta$ is sufficiently large. Countour integrals of analytic functions, in both scalar and matrix case, in the complex plane are easy to evaluate by numerical quadrature like the Trapezoidal rule. In particular, the contour integral in Equation \ref{eq: contour_int_eAt} can be discretized  as \cite{Fang2023timemarchingbased}
\begin{equation}
\label{eq: contour_int_eAt discretized}
    f_K(A) = \frac{1}{K} \sum_{k = 0}^{K-1} e^{z_k} z_k (z_k I_n - A)^{-1} \hspace{.5 cm} \textrm{for} \ \ z_k = \beta e^{i 2\pi k/K}
\end{equation}
Equation \ref{eq: contour_int_eAt discretized} can be implemented through a combination of methods consist of matrix inversion \cite{Gilyn2018QuantumSV} and LCU \cite{Child_LCU, Low2018HamiltonianSim_LCU}. It should be noted that equation \ref{eq: contour_int_eAt discretized} converges exponentially since $e^z$ is an analytic function \cite{davis1959numerical, davis2014methods}. This implies that $K$ will be manageable, and the quantum circuit implementation of \ref{eq: contour_int_eAt discretized} won't have a large LCU overhead. 

\subsection{Overview}
While the quantum approaches discussed above hold promise for delivering quantum speedup in solving complex PDEs, their execution requires large quantum circuits. Currently, achieving such scale is not feasible on existing or near-term hardware due to the absence of error correction. Nevertheless, exploring the capabilities of Noisy-Intermediate Scale Quantum (NISQ) computers remains valuable for two reasons. First, NISQ devices can still offer a quantum advantage for certain problems \cite{IBM_UtilityPaper, Google_Supremacy}. Second, the timeline for the availability of fault-tolerant quantum computers is uncertain. Consequently, considerable effort has been invested in developing algorithms better suited for these machines. This has led to the emergence of a new class of algorithms known as Variational Quantum Algorithms (VQAs) \cite{cerezo2021VQA}, which are hybrid quantum-classical approaches. 

As the name suggests, these algorithms are based on the variational principle \cite{epstein1974variation}. Some parts of the algorithm involve classical computations, while others require quantum computations. The key is that both types of computations—classical and quantum—are efficient, resulting in an overall efficient algorithm and implying quantum speedup. However, quantifying the speedup in variational quantum algorithms is challenging due to their heuristic nature. In many of these algorithms, the classical computer is responsible for minimizing a cost function evaluated on a quantum computer, which can lead to local minima. Nevertheless, they are the most promising quantum algorithms for current and near-term quantum computers. 

The Variational Quantum Eigensolver (VQE) \cite{OriginalVQE, VQE_Review}, the most widely recognized algorithm in this class, for example, serves as an alternative method to Quantum Phase Estimation (QPE) for calculating the electronic ground state energy in challenging quantum chemistry applications \cite{cao2019_qChemqComp}. VQE have demonstrated its capability to accurately predict electronic ground state energy for small molecular systems \cite{kandala2017_IBM_VQE} using NISQ hardware. The Variational Quantum Linear Solver (VQLS) \cite{Carlos_VQLS, Huang2021NearTermQuantumAlgorithms}, building upon VQE, is designed to solve linear systems of equations in a variational manner. Much like VQE offers an alternative quantum approach to compute ground state energy, VQLS provides an alternative perspective to quantum linear system solver algorithms, such as HHL and QSVT, addressing the quantum version of $Ax=b$. Despite their promising features, both VQE and VQLS face challenges, including sensitivity to noise, errors, and convergence issues \cite{Anschuetz2022_VQEChallenges}. Furthermore, when considering the application of VQLS to solve the linear system of equations $\mathcal{A}x = b$ derived from implementing a time integrator for time-dependent PDEs as in Equations \ref{eq:linear_time_dependent_PDE} or \ref{eq:Nonlinear_time_dependent_PDE}, the quantum resources needed extend beyond the NISQ regime. This arises from consolidating the problem at each time step into a single, large linear system of equations (see Equation \ref{eq:FeynmanClock-ForwardEuler}). Additionally, due to the Courant-Friedrichs-Lewy (CFL) condition, which imposes constraints on the time step size, a large number of time steps can lead to an incredibly large linear system of equations requiring more qubits than those available in NISQ hardware.

In this paper, we explore an alternative variational quantum approach to solve the Maxwell’s equations (Equation \ref{eq: complete Maxwell}), namely the Variational Quantum Imaginary Time Evolution (VarQITE) \cite{yuan2019_varQITE_theory, mcardle2019variational}, which is an approximation of the Quantum Imaginary Time Evolution (QITE) \cite{MottaQITE}. While the QITE circuit depth grows linearly with the number of imaginary time steps, making it expensive for NISQ devices, VarQITE maintains a fixed circuit depth along the imaginary-time path, and hence more suitable for the NISQ regime. The structure of this paper is as follows: In Section \ref{sec: Governing Equations}, we provide an overview of Maxwell's equations and their significance in science and engineering applications. We also formulate the specific version of Maxwell's equations that we explore using VarQITE. Section \ref{sec: Background Theory} reviews the background theory of quantum computing, including common notations and concepts. We delve into the theory of VarQITE and how it can be leveraged to solve PDEs. Section \ref{sec: Numerical Approach} presents our numerical approaches, such as discretization schemes, ansatz selections, mesh sizes, and error metrics. In Section \ref{sec: results}, we present our numerical results. Finally, Section \ref{sec: Conclusion} discusses our findings in detail and explores potential avenues for further research and development.

\section{Governing Equations}
\label{sec: Governing Equations}
\subsection{The Maxwell's Equations}
The celebrated Maxwell equations describe the time evolution of electromagnetic fields and waves \cite{Ball2006_MaxwellEq_Theory}. They have widespread importance in the aerospace industry due to their ability to capture the evolution of a very large body of phenomena. Some of these applications are 1. Hypersonic flight \cite{Gaitonde2006MagnetohydrodynamicEP, gaitonde2008highspeed} (which frequently involves the formation of weakly-ionized plasma), 2. Lightning strike effects and resulting damage on composite materials of aircraft \cite{michael2019multiphysics}, 3. Wiring and wiring bundle effects and shielding from stray signals, 4. Radio communications and blackout \cite{Shang2001RadioBlackout}, 5. Electric space propulsion (including arcjet thrusters, ion thrusters, Hall effect thrusters, magnetoplasmadynamic thrusters, etc.) \cite{MHD_Propulsion}, 5. High-energy applications (including nuclear fission and fusion applications) \cite{otin2019computational}, 6. Electrochemical materials behavior (for example, galvanic corrosion action occurs as a direct result of electric behavior, and often requires an electrostatic numerical solution in modeling practices). 

The Maxwell's equation consist of eight partial differential equations \cite{Ball2006_MaxwellEq_Theory, Wolski2011TheoryOE},

\begin{align}
\label{eq: complete Maxwell}
    &\nabla \cdot \mathbf{B} = 0 \\
    &\nabla \cdot \mathbf{D} = \varrho_c \\
    &\frac{\partial \mathbf{B}}{\partial t} + \nabla \times \mathbf{E} = 0\\
    &\frac{\partial \mathbf{D}}{\partial t} + \mathbf{j} = \nabla \times \mathbf{H}
\end{align}
which represent the magnetic divergence constraint, Gauss’s law, Faraday’s law, and Ampere’s law, respectively. Here, $\mathbf{E} = (E_x, E_y, E_z)$ and $\mathbf{H} = (H_x, H_y, H_z)$ are the electric and magnetic field intensities property, $\mathbf{D} = (D_x, D_y, D_z)$ and $\mathbf{B} = (B_x, B_y, B_z)$ are the electric displacement and magnetic induction fields, $\varrho_c$ us the free charge density, and $\mathbf{j}$ is the current density. 

Commonly, the auxiliary fields $\mathbf{H}$ and $\mathbf{D}$ are related to the electric and magnetic fields, $\mathbf{E}$ and $\mathbf{B}$, via the simple approximations $\mathbf{D} = \overleftrightarrow{\overleftrightarrow{\epsilon}} \cdot \mathbf{E} + \mathbf{P}$ and $H = \overleftrightarrow{\overleftrightarrow{\mu}}^{-1} \cdot \mathbf{B} + \mathbf{M}$ where $\overleftrightarrow{\overleftrightarrow{\epsilon}}$ and $\overleftrightarrow{\overleftrightarrow{\mu}}^{-1} $ are the permittivity and permeability tensors, respectively, and $\mathbf{P}$ and $\mathbf{M}$ are the polarization and magnetization vectors, respectively. These properties are material properties of the medium through which the fields occur. In many cases, the permeability and permeability tensors may be approximated as scalars in isotropic media, and often may be further approximated as their vacuum values, $\epsilon_0$ and $\mu_0$, which are the permittivity and permeability of free space, respectively, and $c_0$ is the speed of light in vacuum (related via the Weber relation, $c_0^2 = (\epsilon_0 \mu_0 )^{-1}$).

Thus, in linear, isotropic media, the Maxwell equations are frequently written as:
\begin{align}
    &\nabla \cdot \mathbf{B} = 0 \\
    &\nabla \cdot \mathbf{E} = \varrho_c/\epsilon_0 \\
    &\frac{\partial \mathbf{B}}{\partial t} + \nabla \times \mathbf{E} = 0\\
    &\frac{1}{c_0^2}\frac{\partial \mathbf{E}}{\partial t} + \mu_0\mathbf{j} = \nabla \times \mathbf{B}
\end{align}
Any physical electromagnetic solution must satisfy the above Maxwell equations. One common complication in numerical solutions of these equations is that the divergence constraints (the Gauss laws for the electric and magnetic fields) is challenging to satisfy \cite{Brackbill1980_DivConstraint}. Numerically, the electromagnetic fields are typically solved by discretizing and time-marching the Faraday and Ampere laws, but without insisting that the solutions additionally satisfy the divergence constraints, unphysical solutions can be generated. Several methods are available for attempting to either ensure a divergence-preserving numerical solution, or to clean any divergence error from the numerical solutions at the next time step. This remains a challenging aspect of numerically solving Maxwell’s equations. The electromagnetic field equations may be written explicitly in Cartesian coordinates as 

\begin{align}
  &\frac{\partial {B_x}}{\partial t} + \frac{\partial {E_z}}{\partial y} - \frac{\partial {E_y}}{\partial z} = 0 \\
   &\frac{\partial {B_y}}{\partial t} + \frac{\partial {E_x}}{\partial z} - \frac{\partial {E_y}}{\partial x} = 0 \\
   &\frac{\partial {B_z}}{\partial t} + \frac{\partial {E_y}}{\partial x} - \frac{\partial {E_x}}{\partial y} = 0 \\
  &\frac{1}{c_0^2}\frac{\partial {E_x}}{\partial t} + \mu_0 j_x - \bigg(\frac{\partial B_z}{\partial y} - \frac{\partial B_y}{\partial z} \bigg) = 0 \\
   &\frac{1}{c_0^2}\frac{\partial {E_y}}{\partial t} + \mu_0 j_y - \bigg(\frac{\partial B_x}{\partial z} - \frac{\partial B_y}{\partial x} \bigg) = 0  \\
   &\frac{1}{c_0^2}\frac{\partial {E_z}}{\partial t} + \mu_0 j_z - \bigg(\frac{\partial B_y}{\partial x} - \frac{\partial B_x}{\partial y} \bigg) = 0  \\
\end{align}

\subsection{Problem Formulation}
In one dimension (the $x$ direction), without loss of generality, the above equations can be written compactly in the vector form as 
\begin{equation}
\label{MainMaxwellEquation}
    \dfrac{\partial }{\partial t}\begin{bmatrix}
    B_x \\
    B_y  \\
    B_z  \\
    E_x \\
    E_y  \\
    E_z\\
\end{bmatrix} + 
    \dfrac{\partial }{\partial t}\begin{bmatrix}
    0 \\
    -E_z  \\
    E_y  \\
    0 \\
    c_0^2B_z  \\
    -c_0^2 B_y\\
\end{bmatrix} = 
\begin{bmatrix}
    0 \\
    0  \\
    0 \\
    -j_x/\epsilon_0 \\
    -j_y/\epsilon_0 \\
    -j_z/\epsilon_0\\
\end{bmatrix} 
\end{equation}
This can be written more compactly as 
\begin{equation}
    \dfrac{\partial U}{\partial t} + \dfrac{\partial }{\partial x} F(U) = J
\end{equation}
where $U = (B_x, B_y, B_z, E_x, E_y, E_z)^T$. Setting the spatial derivative operator$\dfrac{\partial}{\partial x} F(U) = H$, $U = |\psi(x,t)\rangle$, and assuming that $J=0$, then we can write equation \ref{MainMaxwellEquation} as 
\begin{equation}
    \dfrac{\partial}{\partial t} |\psi(x,t) \rangle = H(t)|\psi(x,t)\rangle
\end{equation}
where $H(t)=H$ is a time-dependent but not necessarily Hermitian linear operator. It is our goal to simulate the dynamics of the approximation of $|\psi(x,t)\rangle$  through spatial discretization on the quantum computer. The advantage of doing so is that one can increase the number of discretization points (increasing mesh resolution) without having to sacrifice too much in computational resources. This is due to the fact that we can store $|\psi(x,t)\rangle$ in $2^{n-1}$ discretization points, for an arbitrary integer $n$, using $n$-qubit register instead of the $2^n$ classical-bit registers. This is known as the curse of dimensionality that constitutes a major challenge for all classical numerical methods, in both the local methods (grid or mesh methods) or global methods (spectral methods). 

In this paper, we simplify the aforementioned system to illustrate the numerical approach described in \ref{sec: Numerical Approach}. Particularly, we assume that the source term, $J$, on the right-hand side is zero, with no current density present anywhere in the system. Physically, this insists that no current is present, and the resulting electromagnetic fields shall propagate as pure waves through the medium. The introduction of source terms can introduce additional numerical challenges, as it may introduce strong stiffness to the above system of equations, but its inclusion is beyond the scope of exploration in this work, and remains as a future activity to extend this work to include the presence of source terms.

\section{Background Theory}
\label{sec: Background Theory}

\subsection{Quantum Computing}
In this section, we discuss the principles of quantum computing. For additional details, see  \cite{MikeIke}.

The building block of classical computational devices is a bit, a two-state system, which can be $0$ or $1$. The building block of a quantum computer is then a two-state quantum system, known as a qubit. The general state, $|\psi \rangle$, of a qubit can be described by a vector in a 2-dimensional Hilbert space, $\mathbb{C}^2$, that is 
\begin{equation}
    |\psi \rangle = \alpha |0\rangle + \beta |1\rangle  \ \ \ \alpha, \beta \in \mathbb{C} \ \ \ |\alpha|^2 + |\beta|^2 = 1
\end{equation}
where $|0\rangle = (1, 0)^T$, $|1\rangle = (0,1)^T$ are the usual Euclidian basis vectors and they serve as the computational basis in quantum computing. Due to the normalization constraint, the state of the qubit is rather belongs to the smaller 2-dimensional complex projective space $\mathbb{C}\boldsymbol{P^1}$. The space of a qubit is then can be visualized as a 3-dimensional unit sphere due to the Hopf fibration construction. Similarly, the state of an n-qubit system, $|\psi\rangle^n$ belongs to $\mathbb{C}\boldsymbol{P^{2^n-1}}$, and can be written explicitly as 
\begin{equation}
    |\psi\rangle^n = c_0|00\cdots 0\rangle + c_1|00\cdots 1\rangle + \cdots + c_{2^{n-1}} |11\cdots1\rangle \ \ \ \ \sum_i |c_i|^2 = 1
\end{equation}
where $|00\cdots 0\rangle , |00\cdots 1\rangle, \cdots, |11\cdots1\rangle$ are the Euclidean basis vectors of dimension $2^n$. Analogous to classical computer, quantum computer manipulate the quantum state though logical (quantum) gates. However, quantum logical gates are unitary and hence reversible, unlike classical logical gates. Operations on a single qubit can be represent in general through a 2 by 2 unitary matrix, $U_3$, which as three parameters $\theta, \phi$, and $\lambda$. Within the computational basis, this can be written explicitly as   
\begin{equation}
    U_3(\theta, \phi, \lambda) = \begin{bmatrix}
        \cos \dfrac{\theta}{2} & -e^{i\lambda} \sin \dfrac{\theta}{2}\\
        e^{i \phi} \sin \dfrac{\theta}{2} & e^{i(\phi + \lambda)} \cos \dfrac{\theta}{2}
    \end{bmatrix}
\end{equation}
As we will see, in certain cases, the solution or state of the system we are interested will always be real, and hence such general rotation is nonsensical. For instance, if the state of the system is completely real to start, then the following operation ($R_Y$) will maintain this property 
\begin{equation}
    R_Y(\theta) = \begin{bmatrix}
        \cos \dfrac{\theta}{2} & \sin \dfrac{\theta}{2}\\
       \sin \dfrac{\theta}{2} & cos \dfrac{\theta}{2} = e^{-\theta Y}
    \end{bmatrix}
\end{equation}
where $\theta$ is the angle (radians) of rotation about the Y-axis in the Bloch sphere representation. Note that throughout this paper, $X,Y,Z$ are the the typical 2 by 2 Pauli matrices, and $I$ is the 2 by 2 identity matrix. 
\begin{equation}
    X = \begin{bmatrix}
        0 & 1\\ 1 & 0 
    \end{bmatrix} \hspace{1 cm}
    Y = \begin{bmatrix}
        0 & -i\\ i & 0 
    \end{bmatrix} \hspace{1 cm}
    Z = \begin{bmatrix}
        1 & 0\\ 0 & -1
    \end{bmatrix} \hspace{1 cm}
    I = \begin{bmatrix}
        1 & 0\\ 0 & 1
    \end{bmatrix} 
\end{equation}
An arbitrary n-qubit logical gate can be decomposed into a sequence of $U_3$ and one particular two-qubit gate, the controlled-NOT or CNOT (CX) gate, $ CNOT = \begin{bmatrix} I & \mathbf{0}\\ \mathbf{0} & X\end{bmatrix} $ where $\mathbf{0}$ is 2 by 2 zero-matrix. Note that another way to represent this gate is $CNOT = |0\rangle \langle 0 | \otimes I + |1\rangle \langle 1 |$ where $|\cdot \rangle \langle \cdot |$ represents the outer-product operation, and $\otimes$  represents the usual tensor product operation. With this in mind, the controlled-$R_Y$ can be written as 
\begin{equation}
    CR_Y(\theta) = |0\rangle\langle0| \otimes I + |1\rangle\langle1| \otimes R_Y(\theta)
\end{equation}
we will be using this particular gate to design our ansatz (parameterized quantum circuit) to represent the dynamics of the formulated Maxwell’s equation in the next section.

\subsection{Variational Quantum Imaginary Time Evolution}
The imaginary time evolution of a Hamiltonian, denoted as \(H\), to time \(t\) is precisely the operation \(e^{-Ht}\), which may not necessarily be a unitary operator. However, Motta et al. \cite{MottaQITE} proposed a method to approximate this operation with a unitary operator \(U(t)\) at time \(t\). In other words, given an initial state \(|\psi(t_0)\rangle\) at time \(t_0\), we have:
\begin{equation}
    |e^{-Ht} - \alpha(t)U(t)|_2 \leq \epsilon
\end{equation}
for some arbitrarily small \(\epsilon\), and \(\alpha(t) = \langle \psi(t_0) | e^{-2Ht} | \psi(t_0) \rangle\) is a normalization constant. This method is known as Quantum Imaginary Time Evolution (QITE). The key idea is to represent the imaginary time propagator using unitary operators through least-squares fitting. QITE is a powerful technique for determining the ground state of a chemical system described by a Hamiltonian \(H\). This is because, when evolving a state \(|\psi\rangle = \sum_i c_i |E_i\rangle\), where \(|E_0\rangle\) is the ground state and non-degenerate, we have:
 \begin{equation}
     \lim_{t \to \infty} \dfrac{e^{-Ht}}{ \sqrt{\langle \psi | e^{-2Ht} | \psi \rangle } } =  \lim_{t \to \infty} \dfrac{ \sum_i c_i e^{-tE_i} |E_i\rangle }{ \sqrt{\sum_k |c_k|^2 e^{-2t E_k}} } = |E_0\rangle
 \end{equation}
since the amplitudes of the higher states \(|E_i \rangle\), where \(i \geq 1\), decay exponentially in comparison to the amplitude of the ground state. One key feature of QITE is its compatibility with approximating \(e^{-Ht}\) for non-Hermitian \(H\), which can arise in PDE applications.

\begin{figure}
    \centering
    \includegraphics[width=0.8\linewidth]{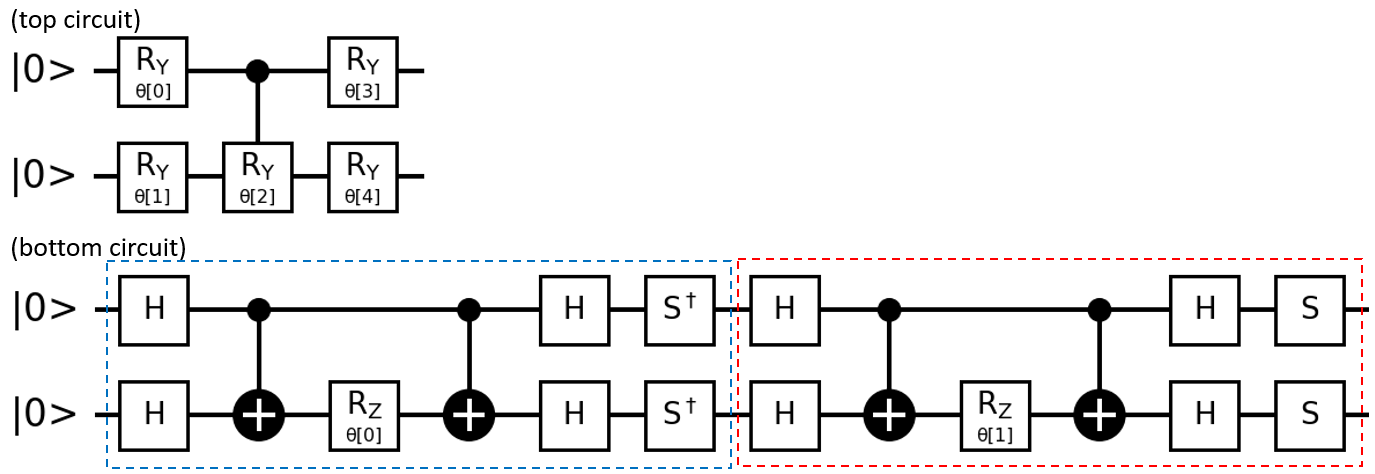}
    \caption{Two examples of the circuit of the form  $U(\boldsymbol{\theta}^t) = \prod_j U_j(\theta_j^t)$. . In particular, the (top circuit) can be written in the product form as $R_Y(\theta_0) \otimes I \cdot I \otimes R_Y(\theta_1) \cdot CR_Y(\theta_2) \otimes I \cdot I \otimes R_Y(\theta_3)$. The (bottom circuit) can be written as $e^{-\theta X\otimes X} \cdot e^{-\theta Y\otimes Y} $   }
    \label{fig:ExampleAnsatz}
\end{figure}

The variational quantum imaginary time evolution (VarQITE) is a hybrid classical-quantum algorithm to approximate the QITE algorithm by mapping it to the ansatz parameters $\boldsymbol{\theta}^t$ by the McLachlan’s variational principle, where $\boldsymbol{\theta}^t = (\theta^t_1, \theta^t_2, \cdots, \theta^t_N) \in \mathbb{R}^N$. In order words, we are approximating the wavefunction $|\psi(t)\rangle $ by the ansatz’s wavefunction $|\phi(t)\rangle = |\phi(\boldsymbol{\theta}^t)\rangle $, that is generated by $U( \boldsymbol{\theta}^t )$ where $U$ is a parametrized quantum circuit or ansatz that is time dependent. In order words, given an initial state $|\psi(t_0)\rangle$, we can approximate the wavefunction at a later time $|\psi(t)\rangle = e^{-Ht}|\psi(t_0)\rangle$ 〉 by an ansatz composed of a series of parameterized quantum gates 
\begin{equation}
\label{Ansatz_Eq}
    U(\boldsymbol{\theta}^t) = \prod_j U_j(\theta_j^t)\\
\end{equation}
\begin{equation}
    |\psi(t) \rangle \approx  |\phi(\boldsymbol{\theta}^t) \rangle = U(\boldsymbol{\theta}^t)|\psi(t_0)\rangle
\end{equation}
where $U_j(\theta)$ is a parametrized quantum gate composed of one parametric rotation gates $e^{-i\theta G}$ where $G = G^\dagger$. Examples of such circuits are shown in Figure \ref{fig:ExampleAnsatz}. VarQITE has been used to solve several PDEs, including the two-dimensional anisotropic heat equation, Poisson equation, and the stochastic differential equation within the Black-Scholes model\cite{Alghassi2022VarQITE}.

Note that we can take the initial state $|\psi(t_0)\rangle$ to be the state $|0\rangle^{\otimes n}$ (as in Figure \ref{fig:ExampleAnsatz} ) without loss of generality. In solving linear PDE, the state $|\phi(\boldsymbol{\theta}^{t_0})\rangle$ represents the initial condition of the ODE system as described by equation \ref{LinearODE} after semi-discretizing the PDE in the spatial domain. If $A$ is independent of time and $\mathbf{b}$ is $\mathbf{0}$ in equation \ref{LinearODE} then its solution is simply $e^{At}u(t_0)$. Hence, the idea of using VarQITE to solve the semi-discretized PDE is to first represent the normalized state \(u(t_0)\) in terms of the wavefunction \(|\psi(t_0)\rangle\). Then, approximate \(|\psi(t_0)\rangle\) with \(|\phi(\boldsymbol{\theta}^{t_0})\rangle\). By understanding how to describe the evolution of \(\boldsymbol{\theta}^t\) as a function of time \(t\), we can construct the normalized state \(u(t)\), \(|\psi(t)\rangle\), through quantum imaginary time evolution. In other words, \(|\psi(t)\rangle = e^{At} |\psi(t_0)\rangle \approx U(\boldsymbol{\theta}^t) |\phi(\boldsymbol{\theta}^{t_0})\rangle\).

In order to determine the optimal parameter $\boldsymbol{\theta}^t$ at each time step $t$, we can minimized the distance between $|\psi(t) $ and $|\phi(\boldsymbol{\theta}^t) \rangle$
\begin{equation}
\label{min_distance_funct}
    \min_{\boldsymbol{\theta}^t} || |\psi(t) \rangle - |\phi(\boldsymbol{\theta}^t) \rangle ||
\end{equation}
Minimizing the distance function in equation \ref{min_distance_funct} is equivalent to solving the McLachlan’s variational principle 
\begin{equation}
\label{McLachlanPrinciple}
\delta ||(\partial_t - H)|\psi(t)\rangle || = 0 
\end{equation}
where $\delta$ represents the infinitesimal variation. By approximating $|\psi(t)\rangle$ with $|\phi(\boldsymbol{\theta}^t) \rangle$, McLachlan’s variational principle yields an Euler-Lagrange-type equation:
\begin{equation}
    \label{eq: Euler-Lagrange}
    \Lambda(t) \partial_t (\boldsymbol{\theta}^t) = C(t) \ \ \forall t
\end{equation}
where 
\begin{equation}
\label{A_ij_coeffs}
    \Lambda_{ij} = \Re\bigg( \frac{\partial \langle \phi(\boldsymbol{\theta}^t))|}{\partial \theta_i^t}  \frac{\partial | \phi(\boldsymbol{\theta}^t)) \rangle}{\partial \theta_j^t} \bigg) = \frac{1}{2}\bigg(\frac{\partial \langle \phi(\boldsymbol{\theta}^t))|}{\partial \theta_i^t}  \frac{\partial | \phi(\boldsymbol{\theta}^t)) \rangle}{\partial \theta_j^t} + \frac{\partial \langle \phi(\boldsymbol{\theta}^t))|}{\partial \theta_j^t}  \frac{\partial | \phi(\boldsymbol{\theta}^t)) \rangle}{\partial \theta_i^t} \bigg)
\end{equation}
\begin{equation}
\label{C_i_coeffs}
    C_i = \Re\bigg( \frac{\partial \langle \phi(\boldsymbol{\theta}^t))|}{\partial \theta_i^t}  H |\phi(\boldsymbol{\theta}^t))\rangle \bigg) = \frac{1}{2}\bigg( \frac{\partial \langle \phi(\boldsymbol{\theta}^t))|}{\partial \theta_i^t}  H |\phi(\boldsymbol{\theta}^t))\rangle  + \langle \phi(\boldsymbol{\theta}^t))| H^\dagger \frac{\partial |\phi(\boldsymbol{\theta}^t)) \rangle}{ \partial \theta_i^t } \bigg)
\end{equation}
the coefficients of the matrix $A$ and the vector $C$ will be determined through measurements on a quantum computer. Once both $\Lambda(t)$ and $C(t)$ are obtained, the time evolution can be computed using a classical computer, for example, the forward Euler method. In this case, we have:
\begin{equation}
\label{eq: VarQITE param update}
    \boldsymbol{\theta}^{t+\Delta t} = \boldsymbol{\theta}^{t} + \Delta t \cdot \partial_t (\boldsymbol{\theta}^{t})  = \boldsymbol{\theta}^{t} + \Delta t \cdot \Lambda^{-1}(t) C(t)
\end{equation}
Here, $\Delta t$ represents a small time step at $t$. Equation \ref{eq: VarQITE param update} highlights a significant difference between VarQITE and Variational Quantum Algorithms (VQAs) like VQLS or VQE. In VQAs, the ansatz parameters are updated through a classical optimizer (e.g., COBYLA, SPSA, SLSQP \cite{VQA_Optimizers_Benchmark}) attempting to optimize the defined cost function. A major issue with VQAs is the optimizer often getting stuck at a local minimum, known as barren plateaus \cite{McClean_2018_Barren_plateaus}. In contrast, for VarQITE, the parameter update simply involves solving the differential equation \ref{eq: Euler-Lagrange}, avoiding the barren plateaus problem encountered by VQAs.

\begin{figure}
    \centering
    \includegraphics[width=0.4\linewidth]{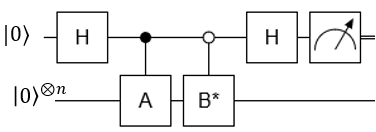}
    \caption{Quantum circuit to compute $^{\otimes n}\langle 0 | AB | 0 \rangle^{\otimes n}$ where $A$ and $B$ are unitary matrices. Note that $B^*$ represents the conjugate transpose of $B$.}
    \label{fig:overlap_circuit]}
\end{figure}

\begin{figure}
    \centering
    \includegraphics[width=0.9\linewidth]{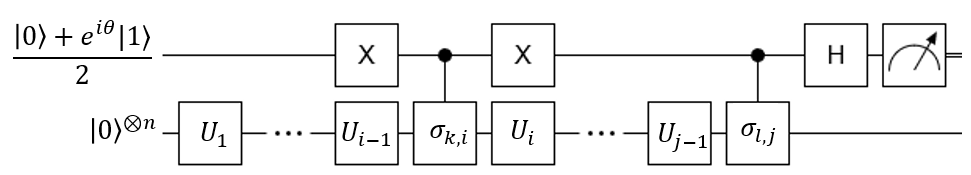}
    \caption{ Quantum circuit to measure $\Re\big( e^{i \theta} \langle0| V_{k,i}^\dagger V_{i,j} | 0 \rangle \big) $. The top qubit is the ancilla qubit.  }
    \label{fig: circuit_A_ij}
\end{figure}

\begin{figure}
    \centering
    \includegraphics[width=0.9\linewidth]{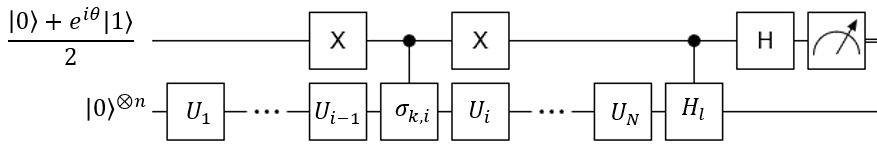}
    \caption{ Quantum circuit to measure $\Re\big( e^{i \theta} \langle0| V_{k,i}^\dagger H_i  | 0 \rangle \big) $. The top qubit is the ancilla qubit. }
    \label{fig: circuit_C_i}
\end{figure}

Unfortunately, $A(t)$ can be ill-conditioned for each $t$, making its inverse problematic. In such scenarios, we will instead perform $\Lambda^{(-1)} (t)$ using either the Moore-Penrose inverse or least squares. Since the elements of the matrix $\Lambda$ and the vector $C$ are determined through measurements on the quantum computer, there will be statistical errors arising from finite sampling. If we choose our ansatz $U$ to have the form of equation \ref{Ansatz_Eq}, then the derivative of each $U_i$ with respect to each $\theta_i$ can be written as:
\begin{equation}
\label{Derivative_of_U}
    \frac{\partial U_i (\theta_i)}{\partial \theta_i} = \sum_k a_{k,i} U_i(\theta_i) \sigma_{k,i}
\end{equation}
where $ a_{k,i}$ is a scalar parameter, and $\sigma_{k,i}$ is a unitary operator. For example, if $U$ is given as (the top circuit in Figure \ref{fig:ExampleAnsatz})
\begin{equation}
    U(\boldsymbol{\theta}) = \underbrace{R_Y(\theta_0) \otimes I}_{U_0(\theta_0)} \cdot \underbrace{I \otimes R_Y(\theta_1) }_{U_1(\theta_1)} \cdot \underbrace{CR_Y(\theta_2)}_{U_2(\theta_2)} \cdot  \underbrace{R_Y(\theta_3) \otimes I}_{U_3(\theta_3)} \cdot \underbrace{I \otimes R_Y(\theta_4) }_{U_4(\theta_4)} 
\end{equation}
then 
\begin{equation}
\label{Derivative_U_example1}
    \frac{\partial U_0}{\partial \theta_0} = -\frac{i}{2} e^{-i \frac{\theta_1 Y}{2} } Y\otimes I = -\overbrace{\frac{i}{2}}^{a_{0,0}} \cdot \overbrace{ R_Y(\theta_0) \otimes I }^{U_0(\theta_0)} \cdot \overbrace{ Y \otimes I }^{\sigma_{0,0}}
\end{equation}
and 
\begin{equation}
\label{Derivative_U_example2}
    \frac{\partial U_2}{\partial \theta_2} = -\frac{i}{2} |1\rangle \langle 1 | \otimes Ye^{-i \frac{\theta_2}{2} Y} = \overbrace{-\frac{i}{4}}^{\alpha_{0,2}} \cdot \overbrace{CR_Y(\theta_2)}^{U_2(\theta_2)} \cdot \overbrace{ I \otimes Y}^{\sigma_{0,2} } + \overbrace{-\frac{i}{4}}^{\alpha_{1,2}} \cdot \overbrace{CR_Y(\theta_2)}^{U_2(\theta_2)} \cdot \overbrace{ Z \otimes Y}^{ \sigma_{1,2} }
\end{equation}
Note that if we are using circuit consist of only single qubit rotation and single controlled-rotation then the expansion of equation \ref{Derivative_of_U} will have at most k=2 as seen in equation \ref{Derivative_U_example1}
 and equation \ref{Derivative_U_example2}. Now, applying equation \ref{Derivative_of_U} to the trial state $|\phi(\boldsymbol{\theta}^t) \rangle = U(\boldsymbol{\theta}^t)|0\rangle^{\otimes n}$ gives 
 \begin{equation}
     \frac{\partial |\phi(\boldsymbol{\theta}^t) \rangle}{ \partial \theta_i} = \sum_k a_{k,i} V_{k,i}|0\rangle^{\otimes n}
 \end{equation}
 with $V_{k,i} = U_N(\theta_N) \cdots U_{i+1}(\theta_{i+1}) \cdot U_i(\theta_i) \sigma_{k,i} \cdot U_{i-1}(\theta_{i-1}) \cdots U_1(\theta_1)$. Thus, the coefficients $A_{i,j}$ and $C_i$ from equations \ref{A_ij_coeffs} and \ref{C_i_coeffs}, respectively, can be rewritten as 
\begin{equation}
\label{A_ij_coeffs_update}
    \Lambda_{ij} = \Re\bigg( \frac{\partial \langle \phi(\boldsymbol{\theta}^t))|}{\partial \theta_i^t}  \frac{\partial | \phi(\boldsymbol{\theta}^t)) \rangle}{\partial \theta_j^t} \bigg) = \Re \bigg( \sum_{k,l} a_{k,l}^* a_{l,j}  ^{\otimes n}\langle 0 | V_{k,i} V_{l,j} | 0 \rangle^{\otimes n} \bigg)
\end{equation}
\begin{equation}
\label{C_i_coeffs_update}
    C_i = \Re\bigg( \frac{\partial \langle \phi(\boldsymbol{\theta}^t))|}{\partial \theta_i^t}  H |\phi(\boldsymbol{\theta}^t))\rangle \bigg) = \Re \bigg(  \sum_{k,l} a_{k,l}^* a_{l,j}   ^{\otimes n}\langle 0| V_{k,i} H_l | 0 \rangle^{\otimes n}  \bigg)
\end{equation}
where \(c_l\) are the coefficients of the expansion of the Hamiltonian, \(H = \sum_i c_i H_i\), with \(H_i\) being unitary operators. For instance, they can be Pauli strings, i.e., \(H_i \in \{I, X, Y, Z\}^{\otimes n}\). The coefficients of the matrix A and vector C are now readily to be evaluated through measurements on a quantum computer using the fact that given two unitary matrices $ A$ and $B$, we can obtain $^{\otimes n}\langle 0 | AB | 0 \rangle^{\otimes n}$  using the circuit shown in Figure \ref{fig:overlap_circuit]}. Based on this circuit structure, we can construct the quantum circuit to extract each of the coefficient $\Lambda_{i,j}$ and $C_i$. See Figure \ref{fig: circuit_A_ij} and \ref{fig: circuit_C_i}.

 In order to extract the coefficients of the matrix \( \Lambda \) and vector \(C\), the circuits in \ref{fig: circuit_A_ij} and \ref{fig: circuit_C_i} must be executed many times to obtain a sufficient number of statistical samples for evaluation. More precisely, for each coefficient, the error in the estimate scales as \(O(1/\sqrt{M})\), where \(M\) is the number of samples due to Chebyshev's inequality \cite{StatsAnalysis}. The size of the matrix \(\Lambda\) and vector \(C\) is determined by the number of free parameters in the ansatz, which, in turn, determines the number of quantum circuits to be executed at each time step \(t_i\). For an ansatz with \(d\) number of free parameters, there will be \(d^2\) and \(d\) quantum circuits needed to be executed to extract the coefficients for the matrix \(\Lambda\) and vector \(C\), respectively. With standard sampling, each circuit will be executed \(O(1/\epsilon^2)\) times to achieve an arbitrary precision \(\epsilon\) for these coefficients \cite{VQE_Measurements_DaveWecker}. Therefore, the number of queries we need to make to the quantum computer for the overall VarQITE process scales as \(O\left(\frac{td^2}{\Delta t \epsilon^2}\right)\), where \(t\) is the total simulation time, \(\Delta t\) is the time step size, \(\epsilon\) is the precision of the evaluation of the coefficients in the matrix \(\Lambda\) and vector \(C\), and \(d\) is the number of free parameters in the ansatz.

\section{Numerical Method }
\label{sec: Numerical Approach}

\subsection{Discretization}
\label{sec: Discretization}
Regardless of whether a classical or quantum numerical method is used, numerical discretization is necessary to transform the system of partial differential equations (equation \ref{MainMaxwellEquation}) into a form suitable for numerical solution on a computing platform. In the approach explored in this paper, this step involves reducing the system of partial differential equations into an effective qubit Hamiltonian, which must be applied and evolved using the variational imaginary time evolution method.

Various discretization schemes and choices are available. Mesh-based methods include finite difference, finite element, finite volume, and finite integration approaches \cite{hoffmannV1, hoffmannV2, hoffmannV3}. Finite difference reduces the system by approximating the partial derivatives in the equations; finite element approaches instead approximate a weak form of the solution to the system. Finite volume and integration approaches adopt an equivalent integral form of the system and discretize this integral form. More exotic non-mesh-based approaches, including lattice Boltzmann and smoothed particle hydrodynamics approaches, as well as spectral methods, also exist. Any choice of discretization introduces a degree of discretization error into the achieved solution.

In this paper, we target a common representation of simulating electromagnetism through the use of a finite-difference time domain (FDTD) discretization. This discretizes the system by assuming a Cartesian grid of $N$ uniformly-spaced points in a grid. The solution vector is stored at each node in the grid, $u_i, i \in \{1,2,\cdot, N\}$. The spatial partial derivatives are approximated via a central difference, which yields a second-order-accurate calculation \cite{LeVequeFDM}, and the temporal derivatives are approximated via a first-order forward-Euler discretization:
\begin{equation}
    \frac{\partial u_i^n}{\partial x} \approx \frac{u_{i+1}^n - u_{i-1}^n}{2\Delta x}, \quad \frac{\partial u_i^n}{\partial t} \approx \frac{u_i^{n+1} - u_i^n}{\Delta t}
\end{equation}
We apply this discretization to the Maxwell system of equations. In the below, we reduce the system by dropping the $x$ components of the electric and magnetic field; the divergence equations in 1D can be satisfied exactly in these cases and are not dealt with here, although in 2D or higher, some divergence-preserving method must be introduced to suppress the spread of divergence error \cite{ThompsonIEEE}. We further assume no presence of current density, $J=0$.

Starting with the governing system of equations, we have
\begin{equation}
    \label{eq: governing system of equations}
    \frac{\partial U}{\partial t} + \frac{\partial}{\partial x} F(U) = 0
\end{equation}
where \( U = \{B_y, B_z, E_y, E_z\} \) and \( F = \{-E_z, E_y, c^2B_z, -c^2B_y\} \). We begin by recasting this equation into a quasilinear form,
\begin{equation}
    \frac{\partial U}{\partial t} + A \cdot \frac{\partial U}{\partial x} = 0
\end{equation}
where \( A = \frac{\partial F}{\partial U} \) is the Jacobian matrix, introduced by the chain rule:
\begin{equation}
    A = \begin{bmatrix}
    0 & 0 & 0 & -1 \\
    0 & 0 & 1 & 0 \\
    0 & c^2 & 0 & 0 \\
    -c^2 & 0 & 0 & 0
\end{bmatrix}
\end{equation}
We may now discretize the system into the form of a linear system of equations using the central and forward-Euler approximations above, which results in the following system of equations:
\begin{equation}
    u_i^{n+1} = u_i^n - \frac{\Delta t}{2\Delta x} \cdot A \cdot \{u_{i+1}^n - u_{i-1}^n\}
\end{equation}

The full system may be expressed as a fully assembled linear system with a matrix \(H \in \mathbb{R}^{4N \times 4N}\) (four scalar variables of \(u_i^n\) for each node \(i\)) for a full solution vector \(U = \{u_1^n, u_2^n, \ldots, u_N^n\} \in \mathbb{R}^{4N}\). Note that the memory scaling on a quantum computer is much more graceful than that of a classical computer. In particular, the number of qubits required to store a vector of dimension $2^N$ is $\log(N)$, representing an exponential saving compared to classical bits. The full evolution of the system of equations in discretized form can be expressed as follows:
\begin{equation}
    U^{n+1} = U^n - \Delta t \cdot H \cdot U^n
\end{equation}
where 
\begin{equation}
\label{ProblemHamiltonian}
    H = \frac{A}{2\Delta x} \cdot \{H_{B_y} + H_{B_z} + H_{E_y} + H_{E_z}\}
\end{equation}
with \(H_{B_y}\), \(H_{B_z}\), \(H_{E_y}\), \(H_{E_z}\) are simple binary matrices assembled from the ordering of the mesh nodes to indicate the positions of \(u_{i+1}\) and \(u_{i-1}\) relative to \(u_i\) for each scalar component.

In the following, we denote $H$ (equation \ref{ProblemHamiltonian}) to represent the Hamiltonian for the variational quantum imaginary time evolution (VarQITE) approach. We will compare the solutions generated by VarQITE with those from the classically-solved FDTD case using the forward-Euler time-marching scheme. It's worth noting that alternative discretization schemes are possible here and could replace the above approach. Moreover, higher-order accuracy could be achieved by implementing larger-stencil schemes, compact differences, or by resorting to other base schemes in the finite element or finite volume families to improve the degree of discretization error introduced. Future work will continue to explore the application and characterization of other discretization approaches and their associated error for use with the VarQITE method.

\subsection{Ansatz}
\label{sec:Ansatz}
The success of variational quantum algorithms like VQE or VQLS hinges on a critical factor: selecting the appropriate Ansatz, represented by a parameterized quantum circuit $U(\theta)$, for training. The same principle holds for VarQITE. It is imperative to choose an Ansatz that strikes a balance between large expressibility and a concise circuit depth. Expressibility, in this context, gauges how uniformly the chosen Ansatz explores the unitary space \cite{holmes2022Expressibility}, a pivotal property for accurate approximations to the solution. One of the most renowned Ansatz families is the Hardware Efficient Ansatz (HEA). HEA's primary objective is to alleviate gates overhead, a common challenge during the compilation of an abstract quantum circuit into a sequence of native gates for a specific hardware and fixed qubit topology. By aligning with the hardware capabilities of a particular quantum computer, HEA aims to effectively harness the computational power of NISQ devices. One drawback of HEA-type Ansatz in VQAs is that they suffer from a training issue, also known as the barren plateaus problem \cite{McClean_2018_Barren_plateaus}. However, since varQITE does not require a training process like that of VQAs; instead, the parameters are updated through solving equation \ref{eq: Euler-Lagrange}. Therefore, HEA-type Ansatz is an ideal candidate.

\begin{figure}[h]
    \centering
    \includegraphics[width=0.6\linewidth]{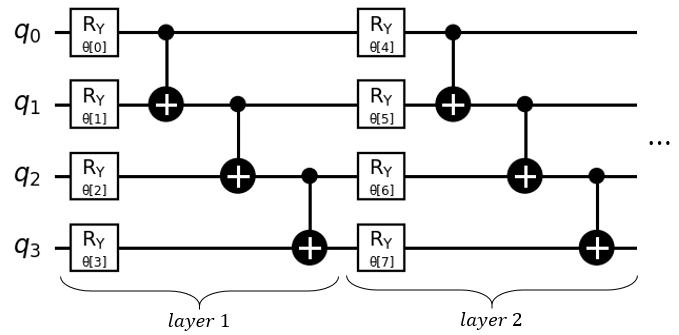}
    \caption{The TwoLocal-Ry-LinearEntanglement ansatz for 4-qubit system. }
    \label{fig:TwoLocal-Ry-LinearEntanglement}
\end{figure}

In this manuscript, we explore several classes of Hardware Efficient Ansatz (HEA). The first class we examine is a parameterized circuit with single-qubit rotational gates \(R_Y\) acting on all qubits, followed by a sequence of CNOT gates acting on neighboring qubits. We refer to this ansatz class as 'TwoLocal-Ry-LinearEntanglement.' See Figure \ref{fig:TwoLocal-Ry-LinearEntanglement} for a 4-qubit example. One notable advantage of this ansatz is its hardware-friendly structure, as \(R_Y\) rotational gates and neighboring qubit CNOT gates can be easily implemented across most quantum hardware platforms (e.g., IBM Quantum, Rigetti, etc.). This results in a shorter compiled circuit, the actual circuit that will be executed by the quantum hardware, ensuring better experimental results due to less noise. Furthermore, this ansatz, particularly the \(R_Y\) gate, was chosen because it ensures that the wavefunction generated through it will always have real-valued coefficients, aligning with our expectations from the formulated Maxwell’s equations.

\begin{figure}
    \centering
    \includegraphics[width=0.85\linewidth]{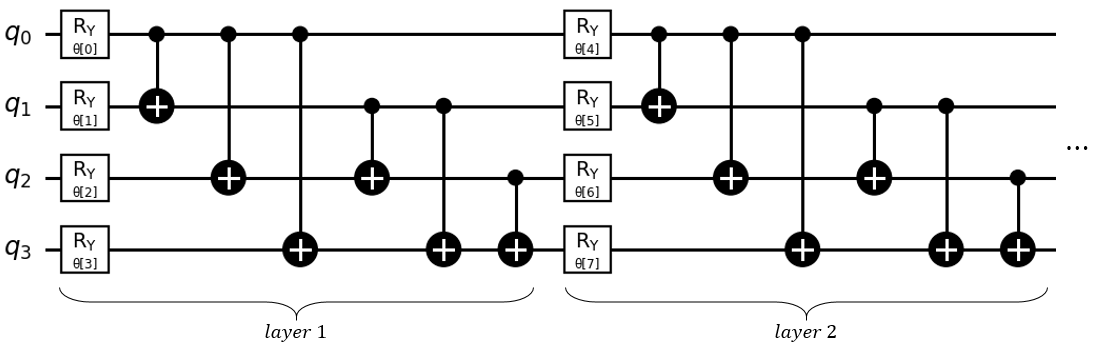}
    \caption{The TwoLocal-Ry-FullEntanglement ansatz for 4-qubit system.}
    \label{fig:TwoLocal-Ry-FullEntanglement}
\end{figure}

In terms of representing this ansatz in the form of \(U(\theta) = \prod_j U_j(\theta)\) with \(U_j(\theta)\) being a parametrized quantum gate composed of one parametric rotation gate, just note that the CNOT gate can also be written as \(CNOT = e^{-i \pi/4(I \otimes I - Z \otimes I)(I \otimes I - I \otimes X)}\). This now allows us to use the circuits in \ref{fig: circuit_A_ij} and \ref{fig: circuit_C_i} to extract the coefficients for the matrix \(A\) and vector \(C\), respectively, to perform our angles update through McLachlan’s principle. We can increase the complexity of this ansatz by having all the qubits interact with one another through different combinations of CNOT gates rather than just neighboring qubits. We call this 'TwoLocal-Ry-FullEntanglement. See Figure \ref{fig:TwoLocal-Ry-FullEntanglement} for a 4-qubit system example.

We can increase the complexity of TwoLocal-Ry-LinearEntanglement and TwoLocal-Ry-FullEntanglement by replacing the CNOT gate with a controlled-rotational gate. However, note that if we replace CNOT with the \(CR_X(\theta)\) gate, then we run the risk of adding complex-valued coefficients to our wavefunction. Hence, we will use the \(CR_Y(\theta)\) as the controlled-rotational gate instead. We will call these two ansatz classes TwoLocal-RyCRy-LinearEntanglement and TwoLocal-RyCRy-FullEntanglement, respectively.
\begin{figure}
    \centering
    \includegraphics[width=0.6\linewidth]{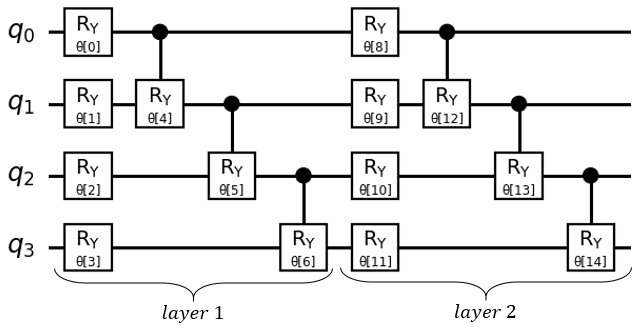}
    \caption{The TwoLocal-RyCRy-LinearEntanglement ansatz for 4-qubit system.}
    \label{fig:TwoLocal-RyCRy-LinearEntanglement}
\end{figure}

For completeness, we can write down an example circuit to measure one of the coefficients of the matrix \(A\) using the ansatz from \ref{fig:TwoLocal-RyCRy-LinearEntanglement}. Without loss of generality, we will measure \(A_{1,10}\).
\begin{figure}
    \centering
    \includegraphics[width=0.9\linewidth]{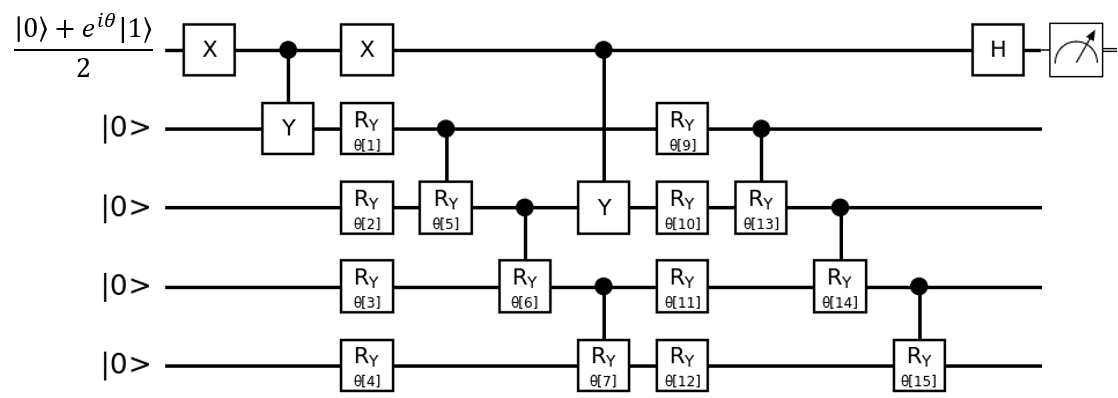}
    \caption{Quantum circuit to extract the coefficient $A_{1.,10}$ using ansatz in Figure \ref{fig:TwoLocal-RyCRy-LinearEntanglement} using equation \ref{A_ij_coeffs_update}}.
    \label{fig:Measuring_A110_for_TwoLocal-RyCRy-LinearEntanglement}
\end{figure}

\subsection{Initial Condition}

In this work, we will impose a Gaussian distribution for the \(B_Z\) component and a constant zero function for \(E_Y\), \(E_Z\), and \(B_Y\) components for our initial conditions, and will perform dynamics on this initialization.
\begin{figure}
    \centering
    \includegraphics[width=0.7\linewidth]{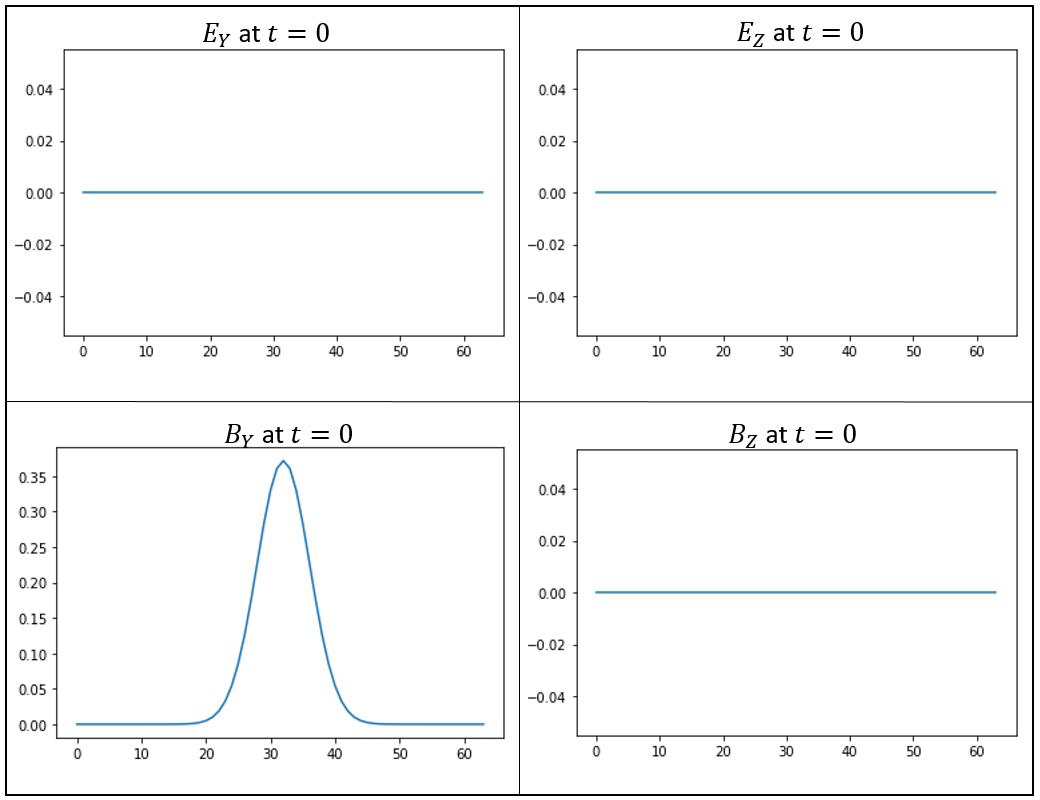}
    \caption{Assumed initial conditions for Equation \ref{MainMaxwellEquation}. }
    \label{fig:Initital_Conditions}
\end{figure}

This problem captures the salient elements of electromagnetic wave propagation in a vacuum. The physical evolution over time of the initial magnetic pulse in the middle of the domain should cause the split propagation of a left-moving and right-moving magnetic pulse traveling at the speed of light, inducing a corollary set of electric pulses in the \(z\) component traveling in opposite directions.

The Gaussian function will be discretized into the chosen mesh points. We can set up an epsilon error threshold regarding the discretization, \(\epsilon_{\text{discretization}}\), and keep decreasing the mesh size (increasing the number of grid points) until the difference between \(|\psi_{\text{actual}}^{(N_{\text{grids}})}(t_0)\rangle\) and \(|\psi_{\text{actual}}(t_0)\rangle\) is less than epsilon, that is
\begin{equation}
    \left\| |\psi_{\text{actual}}^{(N_{\text{grids}})}(t_0)\rangle - |\psi_{\text{actual}}(t_0)\rangle \right\| < \epsilon_{\text{discretization}}
\end{equation}
where \(|\psi_{\text{actual}}^{(N_{\text{grids}})}(t_0)\rangle\) is the projected initial Gaussian function on a mesh with \(N_{\text{grids}}\) grid points. For each \(|\psi_{\text{actual}}^{(N_{\text{grids}})}(t_0)\rangle\), we can approximate it with \(|\phi_{\text{ansatz}}^{(N_{\text{grids}})}(\theta)\rangle = U(\theta) |0\rangle^{N_{\text{grids}}}\) where \(N_{\text{grids}} = 2^N\) and \(U(\theta)\) is the selected ansatz. Thus, for each selected mesh size, we can choose an ansatz and optimize \(U(\theta)\) to minimize the absolute difference between \(|\phi_{\text{ansatz}}^{(N_{\text{grids}})}(\theta)\rangle\) and \(|\psi_{\text{actual}}^{(N_{\text{grids}})}(t_0)\rangle\). That is, we will need to solve the following minimization problem:
\begin{equation}
\label{eq: initial condition cost function}
    \min_{\theta} \| |\phi_{\text{ansatz}}^{(N_{\text{grids}})}(\theta)\rangle - |\psi_{\text{actual}}^{(N_{\text{grids}})}(t_0)\rangle \|
\end{equation}
With the right chosen \(\epsilon_{\text{discretization}}\) and the convergence of the above minimization problem, \(|\phi_{\text{ansatz}}^{(N_{\text{grids}})}(\theta)\rangle\) will correctly represent the initial condition of our problem—a Gaussian distribution function for this particular study. A possible strategy to perform the above minimization is to define a cost function \(C(\theta)\) as
\begin{equation}
    C(\theta) = 1 - \left\| \langle\phi_{\text{ansatz}}^{(N_{\text{grids}})}(\theta)|\psi_{\text{actual}}^{(N_{\text{grids}})}(t_0)\rangle \right\|^2
\end{equation}
and minimize this cost function. The term \(\left\| \langle\phi_{\text{ansatz}}^{(N_{\text{grids}})}(\theta)|\psi_{\text{actual}}^{(N_{\text{grids}})}(t_0)\rangle \right\|^2\) can be calculated on a quantum computer naively using the SWAP test.

\begin{figure}
    \centering
    \includegraphics[width=0.5 \linewidth]{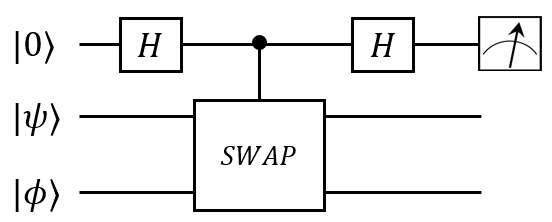}
    \caption{SWAP test to evaluate \(\| \langle \phi_{\text{ansatz}}^{(N_{\text{grids}})}(\theta) | \psi_{\text{actual}}^{(N_{\text{grids}})}(t_0) \rangle \|^2\) to obtain \(C(\theta)\), which will be optimized for \(\theta\) through a classical optimization algorithm.}
    \label{fig:SWAP_Test_Circuit}
\end{figure}

\subsection{Time-average Trace Error Metric }
Rather than only tracking the initial error, we define the error as 
\begin{equation}
    \| \langle \phi_{\text{ansatz}}^{N_{\text{grids}}}(\theta) | - | \psi_{\text{actual}}^{N_{\text{grids}}}(t_0) \rangle \| = \epsilon_0,
\end{equation}
which can propagate if \(|\psi_{\text{actual}}^{N_{\text{grids}}}(t)\rangle\) varies significantly. To address this, we propose an error metric that can keep track of solution quality throughout simulation time, known as the time-average trace error metric, \(\epsilon_{\text{tr}}\). Mathematically, this metric is expressed as
\begin{equation}
   \epsilon_{\text{tr}} = \frac{1}{N} \sum_{i=0}^{N-1} \sqrt{1 - |\langle \psi_{\text{qc}}(t_k) | u^*(t_k) \rangle|^2}, 
\end{equation}
where \(|u^*(t_k)\rangle = \frac{u(t_k)}{\sqrt{\langle u(t_k) | u(t_k) \rangle}}\) is the normalized classical solution vector at time \(t_k\). This enables benchmarking the VarQITE algorithm for classical solvable system sizes. Note that \(u^*(t_k)\) can be computed using methods like Forward Euler, as utilized in our experimental setup. For large systems where the classical solution is intractable, i.e., \(u^*(t_k)\) can't be efficiently prepared with a classical computer, this metric may not be applicable. Nonetheless, we consider it a benchmark candidate metric, serving to test the performance of VarQITE and other quantum PDE solvers against well-known classical solvers.

\subsection{The Algorithm}
To provide a clear summary of the computational workflow for using VarQITE to solve Maxwell’s equations, we have developed the following pseudocode algorithm (see Algorithm \ref{VarQITE Pseudo Algorithm}). The process begins by determining the number of grid points, ${N_{Grids}}$, needed for the discretization of the spatial domain. We must select ${N_{Grids}}$ so that a specified level of discretization error between the actual initial state $|\psi_{actual} (t_0)\rangle$, representing the normalized initial function, and the discretized initial state $|\psi_{actual}^{N_{Grids}} (t_0)\rangle$, denoted as $\epsilon_{discretization}$, is achieved. The calculation to determine ${N_{Grids}}$ for achieving the desired $\epsilon_{discretization}$ can be performed analytically using an error bound. Refer to the literature on numerical quadrature for more details.

The next step in the algorithm is to select an Ansatz (parameterized quantum circuit) family. This step is crucial, as the success of the algorithm heavily depends on choosing the right Ansatz. Once the Ansatz family/structure has been selected, we need to determine the right initial parameters $\boldsymbol{\theta}^0$ so that the Ansatz correctly represents the actual discretized initial state $|\psi_{actual}^{N_{Grids}} (t_0)\rangle$. The calculation for $\boldsymbol{\theta}^0$ can be done through minimizing the cost function described in equation \ref{eq: initial condition cost function}. In certain cases, $\boldsymbol{\theta}^0$ can be determined analytically. It is also possible to construct the initial state $|\psi_{actual}^{N_{Grids}} (t_0)\rangle$ using a quantum circuit that is different from the circuit of the Ansatz, and subsequently, append this circuit to the front of the Ansatz circuit. Many well-behaved functions have efficient quantum circuit implementations \cite{Li_2021_prep_periodic_func, Rattew2021efficient}. This approach would eliminate the initial classical optimization procedure to determine the initial parameters, but the price to pay is a longer circuit depth. For this reason, we elected to prepare the state $|\psi_{actual}^{N_{Grids}} (t_0)\rangle$ by determining the right initial parameters $\boldsymbol{\theta}^0$ through the help of a classical optimizer. See the third step in Algorithm \ref{VarQITE Pseudo Algorithm}. We used the SPSA optimizer from Scipy library \cite{2020SciPy-NMeth}. 

In order to implement VarQITE, the governing equations must undergo semi-discretization using the Method of Lines, as discussed in Section \ref{sec: Introduction}. The discretization scheme employed in this manuscript is detailed in Section \ref{sec: Discretization}. The discretization of the differential operators results in a very sparse matrix representation of the Hamiltonian, denoted as $H$. Subsequently, we decompose $H$ into a linear combination of unitary matrices, i.e., $H = \sum_i c_iH_i$, where each $H_i$ is a unitary operator. Typically, $H_i$ is assumed to be a Pauli string (a tensor product of Pauli matrices and the Identity matrix), represented as $P_i \in { I, X, Y, Z}^{\otimes \log(N)}$, but it is not required. Generally, this decomposition is not efficient, but for sparse matrices, a polynomial scaling decomposition is often feasible.

At this point, we are ready to initiate the VarQITE algorithm after defining the time step size, $\Delta_t$. To ensure numerical stability, $\Delta_t$ needs to be sufficiently small. The choice of $\Delta_t$ significantly influences the total run-time of the overall algorithm, as the number of iterations is given by $N_{iter} = t_{final}/\Delta_t$, where $t_{final}$ is the final time step specified by the user. To advance the initial state forward to the desired $t_{final}$ with VarQITE, we will perform the calculation to extract the matrix $\Lambda$ and the vector $C$ belonging to equation \ref{eq: Euler-Lagrange}, solving it at each iteration step within $N_{iter}$. The construction of the ansatz circuit, as well as the implementation of circuits to extract $\Lambda$ and $C$, was carried out using Qiskit \cite{QiskitCommunity2017}.

\include{pythonlisting}
\begin{algorithm}[H]
\label{VarQITE Pseudo Algorithm}
\SetAlgoLined

initialize ${N_{Grids}}$  such that $$\big| |\psi_{actual}^{N_{Grids}} (t_0)\rangle  - |\psi_{actual} (t_0)\rangle  \big| \leq \epsilon_{discretization} $$ \\

\textbf{select} Ansatz family type (e.g. TwoLocal-RY-LinearEntanglement) of size $N = \log({N_{Grids}})$ qubits

 \While{ $\big| |\phi_{actual}^{N_{Grids}} (\theta)\rangle  - |\psi_{actual}^{N_{Grids}} (t_0)\rangle  \big| \geq \epsilon_{init} $}
 {
 \textbf{increase} Ansatz depth

  $\min_{ \boldsymbol{\theta}^0 } C( \boldsymbol{\theta}^0 )$ where 
  $$ C(\boldsymbol{\theta}^0) =1- \big|\big| \langle\phi_{actual}^{N_{Grids}} (\boldsymbol{\theta}^0)|\psi_{actual}^{N_{Grids}} (t_0)\rangle   \big|\big|^2 $$
 }
 starting with the governing equations as described in equation \ref{eq: governing system of equations}\\
 
 derive the Hamiltonian and discretize it into a sparse matrix $H$ of dimension ${N_{Grids}} \times {N_{Grids}}$ as described in equation \ref{ProblemHamiltonian} \\
 
 decompose the discretized matrix $H$ into the form $H = \sum_i c_iH_i$ where each $H_i$ is a unitary operator\\
 
 set time step size: $\Delta_t$ \\

 set $t =0$, and a fixed $t_{final}$  value\\

\While{ $t < t_{final}$ }
{
Compute $\Lambda$ of equation \ref{eq: Euler-Lagrange} using Equation \ref{A_ij_coeffs}. Each coefficient of the matrix $A$  can be evaluated on the quantum computer using  a quantum circuit that has structure as shown in Figure \ref{fig: circuit_A_ij}. \\ 

Compute $C$ of equation \ref{eq: Euler-Lagrange} using Equation \ref{C_i_coeffs}. Each coefficient of the vector $C$  can be evaluated on the quantum computer using  a quantum circuit that has structure as shown in Figure \ref{fig: circuit_C_i}. 

$\boldsymbol{\theta}^{t + \Delta t} = \boldsymbol{\theta}^t + \Delta_t A^{-1}(t) C(t)$

$t = t + \Delta_t$
} 

\caption{Solving Maxwell's Equations with VarQITE}
\end{algorithm}

\section{Numerical Results}
\label{sec: results}
This section presents the numerical results obtained by applying VarQITE to solve the formulated Maxwell’s equations. We demonstrate our numerical studies by investigating various classes of ansatz with different levels of complexity, achieved by tuning the circuit depth and the number of parameters. Within each ansatz family, we examine the convergence of the numerical solution as a function of the number of parameters and at different mesh sizes or the number of grid points \(N_{\text{grids}}\). The number of qubits (\(N_{\text{qubit}}\)) is indicated in Table \ref{tab:Number of params for Ansatz}.

\begin{figure}
    \centering
    \includegraphics[width=0.55\linewidth]{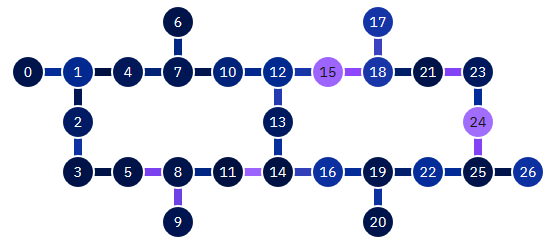}
    \caption{Qubit topology graph of IBM\_Hanoi.}
    \label{fig:IBM_Hanoi}
\end{figure}

The number of parameters scales with respect to circuit layers ($L$), significantly impacting the runtime of the VarQITE algorithm. Here, we explicitly provide the number of parameters used in each of our ansatz families with respect to the number of layers and qubits. In each of the subsection below, we present the actual circuit depth and the compiled circuit depth for different Ansatz types. The compiled circuit depth was computed by decomposing the gates into the native set of gates $S={ I,R_Z (\theta), \sqrt{X},X,CNOT}$ into quantum hardware topology consist of only nearest-neighbors connections like the heavy-hex hardware topology developing by IBM. In particular, we will map out circuit onto a specific IBM’s quantum computer, IBM\_Hanoi, which has the qubit topology shown in Figure \ref{fig:IBM_Hanoi}. It should be noted that the compiled circuit to IBM\_Hanoi was done using a predefined optimization routine in Qiskit, and we selected the highest level of optimization available. By default, Qiskit uses stochastic heuristic algorithms as part of this process, and hence, the circuit depth listed in the tables below will vary slightly on different runs. 

\begin{table}
    \centering
    \begin{tabular}{c|c}
        \hline
         Ansatz Type & Number of Parameters \\
         \hline
         TwoLocal-Ry-LinearEntanglement & $L\cdot N_{\text{qubits}} $\\
         TwoLocal-Ry-FullEntanglement & $L\cdot N_{\text{qubits}}$ \\
         TwoLocal-RyCRy-LinearEntanglement & $(L + N_{\text{qubits}}-1)\cdot N_{\text{qubits}} $\\
         TwoLocal-RyCRy-FullEntanglement & $ \bigg( L + \binom{N_{\text{qubits}}}{2} \bigg) {N_{\text{qubits}}} $\\
         \hline
    \end{tabular}
    \caption{Number of parameters for each ansatz class as a function of number of layers, L, and number of qubits $N_{\text{qubits}}$. See Appendix for explicit values for number of parameters along with circuit depth for each of the ansatz used in this paper.}
    \label{tab:Number of params for Ansatz}
\end{table}

\subsection{TwoLocal-Ry-LinearEntanglement Ansatz }
Here we will outline the results we obtain from the TwoLocal-Ry-LinearEntanglement ansatz  at different mesh (number of grid points) resolution. 

\begin{table}[htb]
\centering
\includegraphics[width=0.9\linewidth]{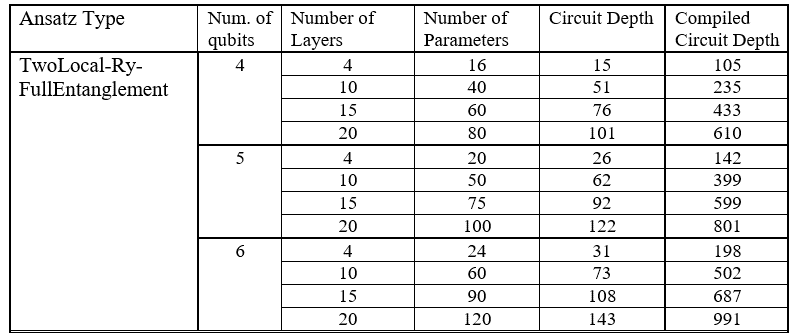}
\label{tab:my table} 
\caption{Variation in the number of parameters, circuit depth, and compiled circuit depth on IBM\_Hanoi for the TwoLocal-Ry-LinearEntanglement ansatz (see Figure \ref{fig:TwoLocal-Ry-LinearEntanglement} ) with different qubits and layers.} 
\end{table}

\subsubsection{16 Grid Points (6-qubit) Analysis}

\begin{figure}
    \centering
    \includegraphics[width=0.9 \linewidth]{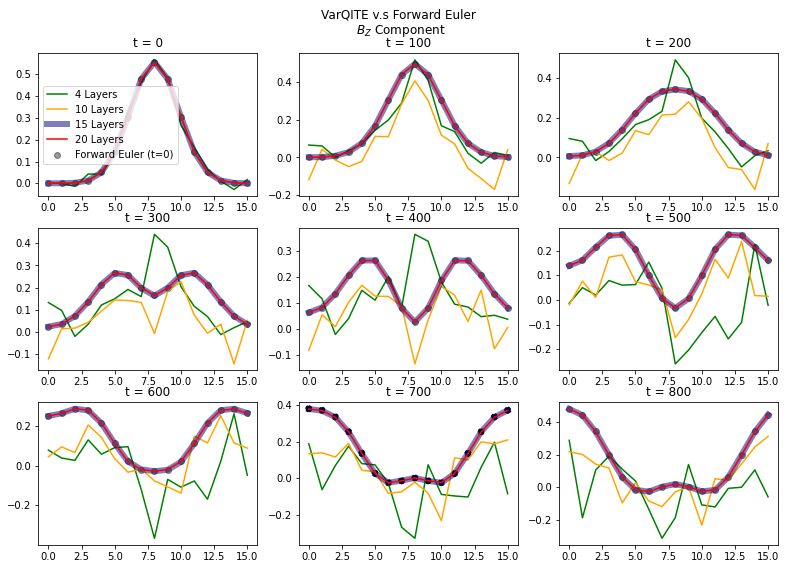}
    \caption{VarQITE simulation under the TwoLocal-Ry-LinearEntanglement Ansatz results in the $B_Z$ component in \ref{MainMaxwellEquation} at different times using a mesh consisting of 16 grid points. The x-axis represents the discretized spatial domain within the internal range of [0,1].}
    \label{fig:TwoLocal-Ry-LinearEntanglement_16Grids_BZ}
\end{figure}

\begin{figure}
    \centering
    \includegraphics[width=0.9 \linewidth]{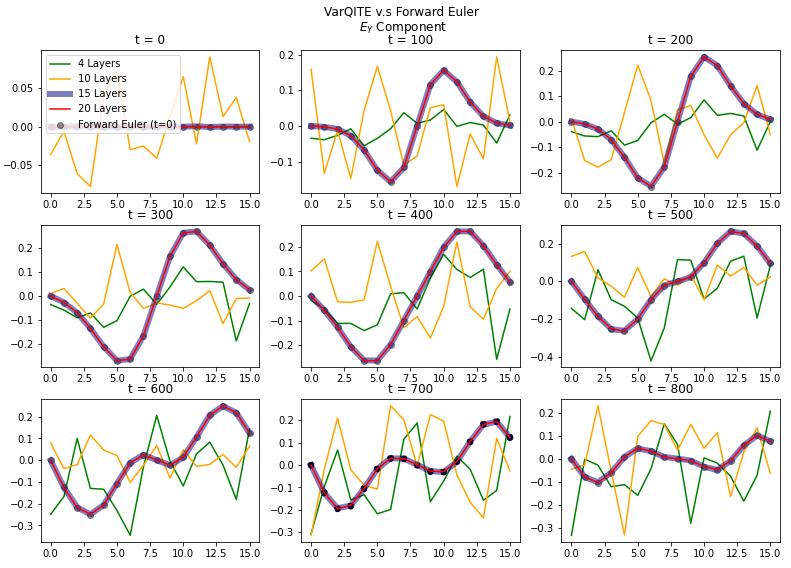}
    \caption{VarQITE simulation under the TwoLocal-Ry-LinearEntanglement Ansatz results in the $E_Y$ component in \ref{MainMaxwellEquation} at different times using a mesh consisting of 16 grid points. The x-axis represents the discretized spatial domain within the internal range of [0,1].  }
    \label{fig:LinearEntanglement_16Grids_EY}
\end{figure}

\begin{figure}
    \centering
    \includegraphics[width=0.6\linewidth]{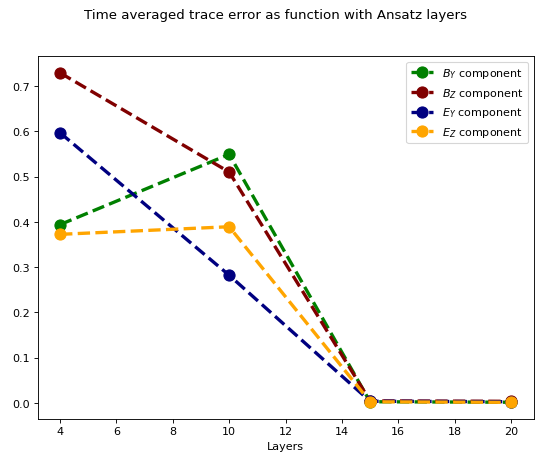}
    \caption{Time-average trace error metric in VarQITE simulation with TwoLocal-Ry-LinearEntanglement Ansatz. Reference solution is obtained from classically-solved FDTD case using the forward-Euler time-marching scheme. Simulation mesh comprises 16 grid points. The x-axis represents the number of layers in the ansatz, while the y-axis represents the measure of error.}
    \label{fig:LinearEntanglement_16Grids_ErrorMetric}
\end{figure}

\clearpage
\subsubsection{32 Grid Points (7-qubit) Analysis }

\begin{figure}[htb]
    \centering
    \includegraphics[width=0.9\linewidth]{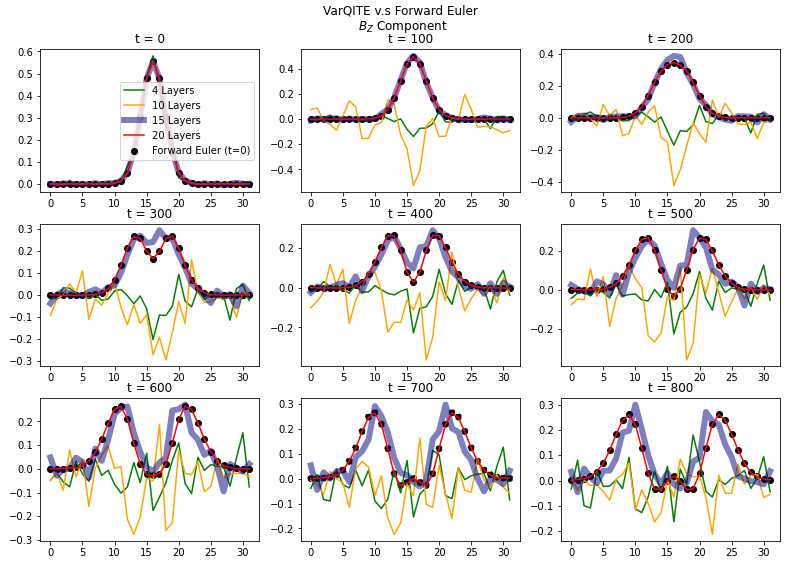}
    \caption{VarQITE simulation under the TwoLocal-Ry-LinearEntanglement Ansatz results in the $B_Z$ component in \ref{MainMaxwellEquation} at different times using a mesh consisting of 32 grid points. The x-axis represents the discretized spatial domain within the internal range of [0,1].  }
    \label{fig:TwoLocal-Ry-LinearEntanglement_BZ_32Grids}
\end{figure}

\begin{figure}
    \centering
    \includegraphics[width=0.9\linewidth]{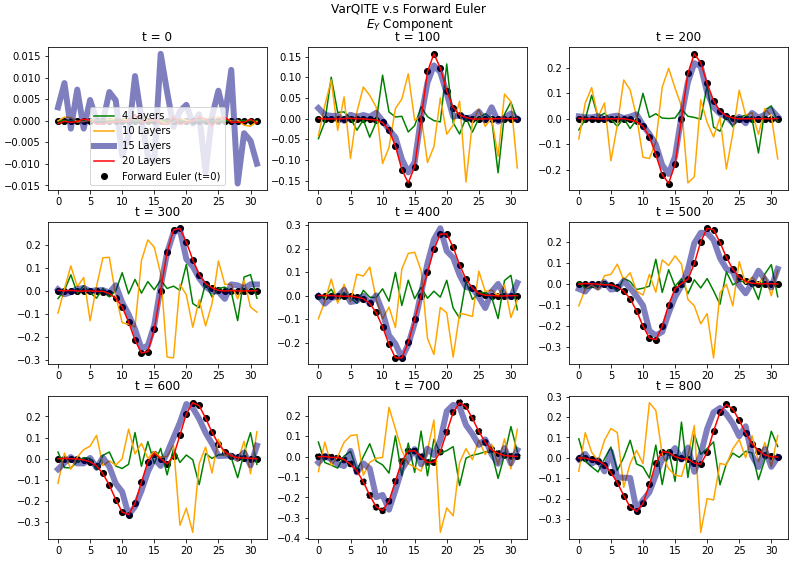}
    \caption{VarQITE simulation under the TwoLocal-Ry-LinearEntanglement Ansatz results in the $E_Y$ component in \ref{MainMaxwellEquation} at different times using a mesh consisting of 32 grid points. The x-axis represents the discretized spatial domain within the internal range of [0,1]. }
    \label{fig:TwoLocal-Ry-LinearEntanglement_EY_32Grids}
\end{figure}

\begin{figure}
    \centering
    \includegraphics[width=0.6\linewidth]{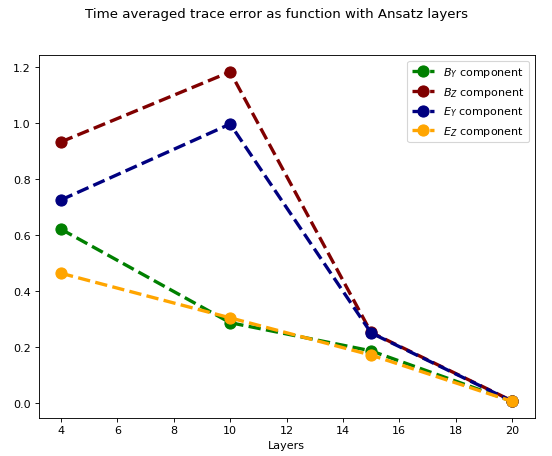}
    \caption{Time-average trace error metric in VarQITE simulation with TwoLocal-Ry-LinearEntanglement Ansatz. Reference solution is obtained from classically-solved FDTD case using the forward-Euler time-marching scheme. Simulation mesh comprises 32 grid points. The x-axis represents the number of layers in the ansatz, while the y-axis represents the measure of error.}
    \label{fig:TwoLocal-Ry-LinearEntanglement Error Metric 32 grids}
\end{figure}

\clearpage
\subsection{TwoLocal-Ry-FullEntanglement Ansatz  }
Here we will outline the results we obtain from the TwoLocal-Ry-LinearEntanglement ansatz  at different mesh (number of grid points) resolution. 

\begin{table}[htb]
\centering
\includegraphics[width=0.9\linewidth]{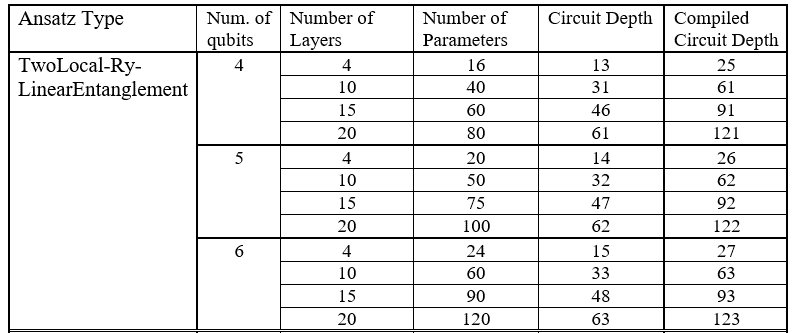}
\label{tab:TwoLocalRYFull_ParamTable} 
\caption{Variation in the number of parameters, circuit depth, and compiled circuit depth on IBM\_Hanoi for the TwoLocal-Ry-FullEntanglement ansatz (see Figure \ref{fig:TwoLocal-Ry-FullEntanglement} ) with different qubits and layers.} 
\end{table}

\subsubsection{16 Grid Points (6-qubit) Analysis}

\begin{figure}[htb]
    \centering
    \includegraphics[width=0.9\linewidth]{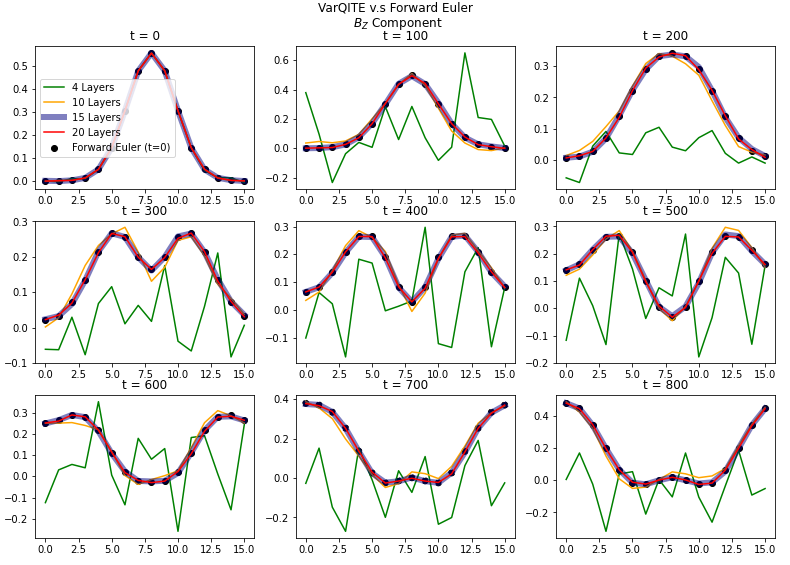}
    \caption{VarQITE simulation under the TwoLocal-Ry-FullEntanglement Ansatz results in the $B_Z$ component in \ref{MainMaxwellEquation} at different times using a mesh consisting of 16 grid points.  The x-axis represents the discretized spatial domain within the internal range of [0,1].}
    \label{fig:TwoLocal-Ry-FullEntanglement_BZ_16Grids}
\end{figure}

\begin{figure}[htb]
    \centering
    \includegraphics[width=0.9\linewidth]{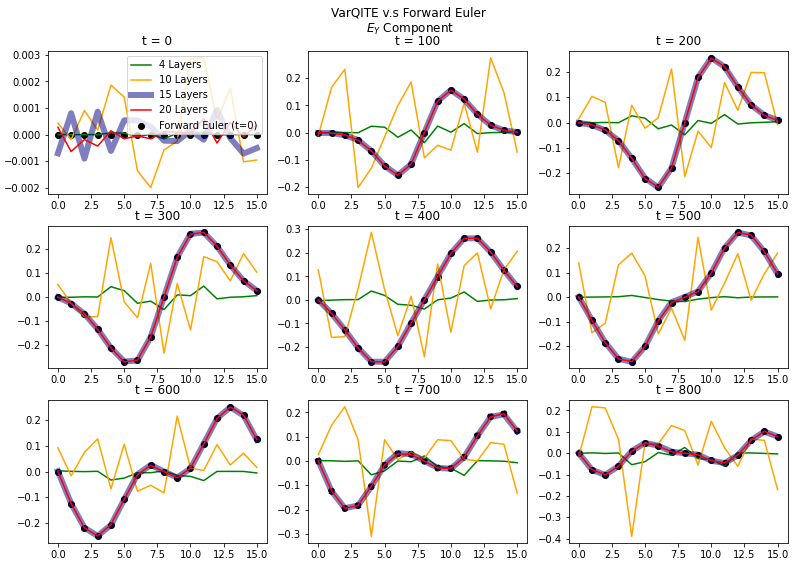}
    \caption{VarQITE simulation under the TwoLocal-Ry-FullEntanglement Ansatz results in the $E_Y$ component in \ref{MainMaxwellEquation} at different times using a mesh consisting of 16 grid points. The x-axis represents the discretized spatial domain within the internal range of [0,1].}
    \label{fig:TwoLocal-Ry-FullEntanglement_EY_16Grids}
\end{figure}

\begin{figure}
    \centering
    \includegraphics[width=0.6\linewidth]{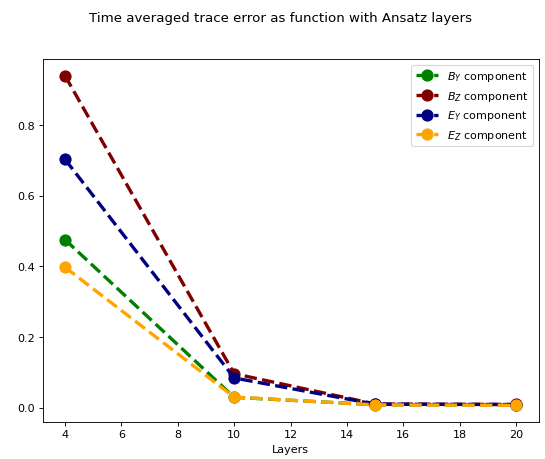}
    \caption{Time-average trace error metric in VarQITE simulation with TwoLocal-Ry-FullEntanglement Ansatz. Reference solution is obtained from classically-solved FDTD case using the forward-Euler time-marching scheme. Simulation mesh comprises 16 grid points. The x-axis represents the number of layers in the ansatz, while the y-axis represents the measure of error.}
    \label{fig:TwoLocal-Ry-FullEntanglement_ErrorMetric_16Grids}
\end{figure}

\clearpage
\subsubsection{32 Grid Points (7-qubit) Analysis}

\begin{figure}[htb]
    \centering
    \includegraphics[width=0.9\linewidth]{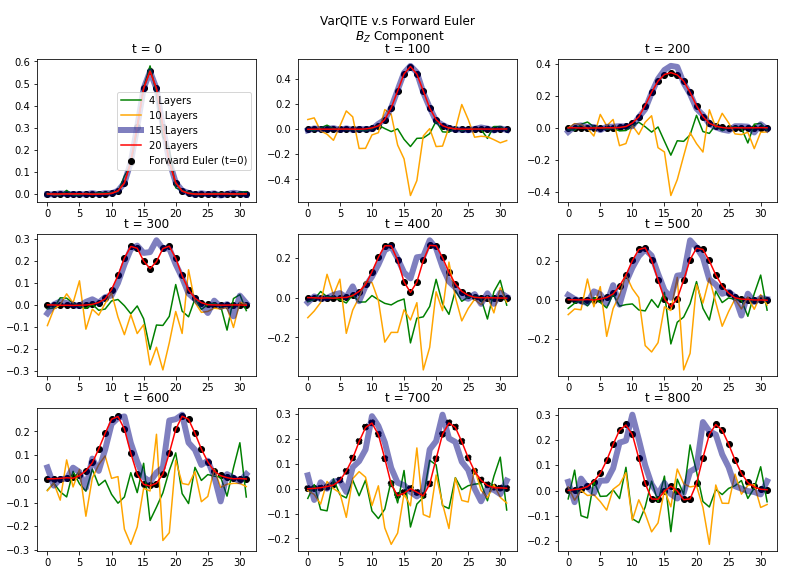}
    \caption{VarQITE simulation under the TwoLocal-Ry-FullEntanglement Ansatz results in the $B_Z$ component in \ref{MainMaxwellEquation} at different times using a mesh consisting of 32 grid points. The x-axis represents the discretized spatial domain within the internal range of [0,1].}
    \label{fig:TwoLocal-Ry-FullEntanglement_BZ_32Grids}
\end{figure}

\begin{figure}[htb]
    \centering
    \includegraphics[width=0.9\linewidth]{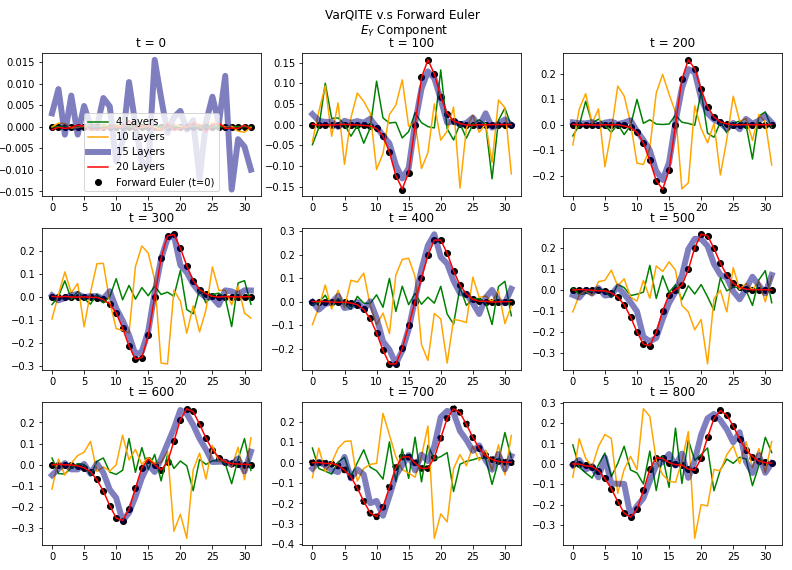}
    \caption{VarQITE simulation under the TwoLocal-Ry-FullEntanglement Ansatz results in the $E_Y$ component in \ref{MainMaxwellEquation} at different times using a mesh consisting of 32 grid points. The x-axis represents the discretized spatial domain within the internal range of [0,1].}
    \label{fig:enter-label}
\end{figure}

\begin{figure}
    \centering
    \includegraphics[width=0.6\linewidth]{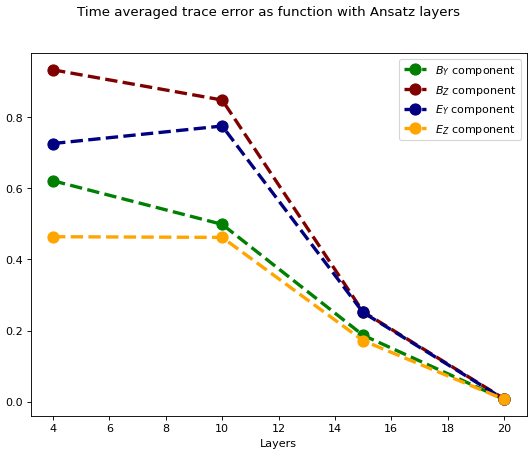}
    \caption{Time-average trace error metric in VarQITE simulation with TwoLocal-Ry-FullEntanglement Ansatz. Reference solution is obtained from classically-solved FDTD case using the forward-Euler time-marching scheme. Simulation mesh comprises 32 grid points. The x-axis represents the number of layers in the ansatz, while the y-axis represents the measure of error.}
    \label{fig:TwoLocal-Ry-FullEntanglement_ErrorMetric_32Grids}
\end{figure}

\section{Conclusion}
\label{sec: Conclusion}
In this paper, we employed the Variational Quantum Imaginary Time Evolution (VarQITE) method to address the simplified Maxwell’s equations (\ref{MainMaxwellEquation}). This approach can be extended to tackle the complete Maxwell’s equations (\ref{eq: complete Maxwell}) as well as numerous other time-dependent differential equations, such as the compressible Navier-Stokes equation. Our findings suggest the feasibility of using VarQITE to solve Maxwell’s equations.  However, we observed that the circuit depth of the ansatz scales exponentially with the number of qubits which corresponded to the number of grid points used in any particular discretization scheme when employing different families of Hardware Efficient Ansatz (HEA), as outlined in Section \ref{sec:Ansatz}. This observation implies that there is no quantum speed-up when using VarQITE to solve Maxwell's equations using HEA as a model to represent the solution space. HEA is often used as a starting point for exploring a variational approach when it is not straightforward to incorporate knowledge about the problem and its solution in a quantum circuit form. However, it is possible to construct a problem-specific Ansatz that offers better scaling than HEA, achieving polynomial depth with respect to the number of qubits instead of exponential.  This is akin to the case of using chemistry-motivated Ansatz to solve for the electronic ground state energy with the VQE algorithm. Nonetheless, this possibility remains an open question for the Maxwell's equation and will be investigated further in our future work.

\section*{Acknowledgments}

We acknowledge Hiromichi Nishimura and Aaila Ali for useful discussions and feedback about the manuscript. We would like to thank the Boeing DC\&N organization, Jay Lowell, and Marna Kagele for creating an environment that made this research/work possible.

\clearpage

\bibliographystyle{iopart-num}
\bibliography{references}

\providecommand{\newblock}{}
\begin{thebibliography}{10}
\expandafter\ifx\csname url\endcsname\relax
  \def\url#1{{\tt #1}}\fi
\expandafter\ifx\csname urlprefix\endcsname\relax\def\urlprefix{URL }\fi
\providecommand{\eprint}[2][]{\url{#2}}

\bibitem{Tinoco1998_BoeingCFD}
Tinoco E~N 1998 The changing role of computational fluid dynamics in aircraft development \urlprefix\url{https://api.semanticscholar.org/CorpusID:111403817}

\bibitem{Johnson2003ThirtyYO}
Johnson F~T, Tinoco E~N and Yu N~J 2003 Thirty years of development and application of cfd at boeing commercial airplanes, seattle \urlprefix\url{https://api.semanticscholar.org/CorpusID:62883377}

\bibitem{Tinoco2007_BoeingCFD}
Tinoco E~N 2007 Cfd uncertainty and validation for commercial aircraft applications \urlprefix\url{https://api.semanticscholar.org/CorpusID:201071086}

\bibitem{Wang2015TowardsHA}
Wang Q 2015 Towards high-fidelity aerospace design in the age of extreme scale supercomputing (invited) \urlprefix\url{https://api.semanticscholar.org/CorpusID:15946019}

\bibitem{jameson2006DNS}
Jameson A and Fatica M 2006 Using computational fluid dynamics for aerodynamics

\bibitem{Reiher2017}
Reiher M, Wiebe N, Svore K~M, Wecker D and Troyer M 2017 {\em Proc Natl Acad Sci U S A\/} {\bf 114} 7555--7560 ISSN 1091-6490 (Electronic), 0027-8424 (Print) \urlprefix\url{https://doi.org/10.1073/pnas.1619152114}

\bibitem{Bauer2020}
Bauer B, Bravyi S, Motta M and Chan G~K~L 2020 {\em Chemical Reviews\/} {\bf 120} 12685--12717 ISSN 0009-2665 \urlprefix\url{https://doi.org/10.1021/acs.chemrev.9b00829}

\bibitem{santagati2023DrugDesign}
Santagati R, Aspuru-Guzik A, Babbush R, Degroote M, Gonzalez L, Kyoseva E, Moll N, Oppel M, Parrish R~M, Rubin N~C, Streif M, Tautermann C~S, Weiss H, Wiebe N and Utschig-Utschig C 2023 {\em arXiv preprint arXiv:2301.04114\/}

\bibitem{HHL_2008}
Harrow A~W, Hassidim A and Lloyd S 2009 {\em Phys. Rev. Lett.\/} {\bf 103}(15) 150502 \urlprefix\url{https://link.aps.org/doi/10.1103/PhysRevLett.103.150502}

\bibitem{CKR}
Childs A~M, Kothari R and Somma R~D 2017 {\em SIAM Journal on Computing\/} {\bf 46} 1920--1950 \urlprefix\url{https://doi.org/10.1137/16M1087072}

\bibitem{Gilyn2018QuantumSV}
Gily{\'e}n A, Su Y, Low G~H and Wiebe N 2018 {\em Proceedings of the 51st Annual ACM SIGACT Symposium on Theory of Computing\/} \urlprefix\url{https://api.semanticscholar.org/CorpusID:46941335}

\bibitem{costa2022optimalLinearSystem}
Costa P~C, An D, Sanders Y~R, Su Y, Babbush R and Berry D~W 2022 {\em PRX Quantum\/} {\bf 3} 040303

\bibitem{an2022optimalLinearSystem}
An D and Lin L 2022 {\em ACM Transactions on Quantum Computing\/} {\bf 3} 1--28

\bibitem{Shor91}
Shor P~W 1997 {\em SIAM Journal on Computing\/} {\bf 26} 1484--1509 (\textit{Preprint} \eprint{https://doi.org/10.1137/S0097539795293172}) \urlprefix\url{https://doi.org/10.1137/S0097539795293172}

\bibitem{Lapworth1}
Lapworth L 2022 {\em Preprint\/} (\textit{Preprint} \eprint{2209.07964})

\bibitem{Lapworth2}
Lapworth L 2022 {\em Preprint\/} (\textit{Preprint} \eprint{2206.00419})

\bibitem{Meng:2023zek}
Meng Z and Yang Y 2023 {\em Phys. Rev. Res.\/} {\bf 5} 033182 (\textit{Preprint} \eprint{2302.09741})

\bibitem{Cao_2013}
Cao Y, Papageorgiou A, Petras I, Traub J and Kais S 2013 {\em New Journal of Physics\/} {\bf 15} 013021 \urlprefix\url{https://dx.doi.org/10.1088/1367-2630/15/1/013021}

\bibitem{Liu_Poisson}
Liu H~L, Wu Y~S, Wan L~C, Pan S~J, Qin S~J, Gao F and Wen Q~Y 2021 {\em Phys. Rev. A\/} {\bf 104}(2) 022418 \urlprefix\url{https://link.aps.org/doi/10.1103/PhysRevA.104.022418}

\bibitem{Wang2020}
Wang S, Wang Z, Li W, Fan L, Wei Z and Gu Y 2020 {\em Quantum Information Processing\/} {\bf 19} 170 ISSN 1573-1332 \urlprefix\url{https://doi.org/10.1007/s11128-020-02669-7}

\bibitem{MOL}
Cutlip M~B and Shacham M 1999 The numerical method of lines for partial differential equations {\em Engineering, Physics\/} \urlprefix\url{https://api.semanticscholar.org/CorpusID:117014395}

\bibitem{Berry_2014}
Berry D~W 2014 {\em Journal of Physics A: Mathematical and Theoretical\/} {\bf 47} 105301 \urlprefix\url{https://dx.doi.org/10.1088/1751-8113/47/10/105301}

\bibitem{Satofuka1987}
Satofuka N, Morinishi K and Nishida Y 1987 {\em Numerical Solution of Two-Dimensional Compressible Navier-Stokes Equations Using Rational Runge-Kutta Method\/} (Wiesbaden: Vieweg+Teubner Verlag) pp 201--218 ISBN 978-3-322-87873-1 \urlprefix\url{https://doi.org/10.1007/978-3-322-87873-1_12}

\bibitem{Oymak1996}
Oymak O and Selcuk N 1996 {\em International Journal for Numerical Methods in Fluids\/} {\bf 23} 455--466

\bibitem{Gaitan2020}
Gaitan F 2020 {\em npj Quantum Information\/} {\bf 6} 61 ISSN 2056-6387 \urlprefix\url{https://doi.org/10.1038/s41534-020-00291-0}

\bibitem{Oz2021}
Oz F, Vuppala R~K~S~S, Kara K and Gaitan F 2021 {\em Quantum Information Processing\/} {\bf 21} 30 ISSN 1573-1332 \urlprefix\url{https://doi.org/10.1007/s11128-021-03391-8}

\bibitem{Liu2021}
Liu J~P, Kolden H~O, Krovi H~K, Loureiro N~F, Trivisa K {\em et~al.\/} 2021 {\em Proceedings of the National Academy of Sciences of the United States of America\/} {\bf 118} e2107884118 ISSN 1091-6490

\bibitem{CarlemanEmbedding}
Kowalski K and Steeb W~H 1991 {\em Nonlinear Dynamical Systems and Carleman Linearization\/} n/a ed (Singapore: World Scientific) ISBN 9810205872

\bibitem{KoopmanTheory}
Brunton S~L, Budi\v{s}i\'{c} M, Kaiser E and Kutz J~N 2022 {\em SIAM Review\/} {\bf 64} 229--340 (\textit{Preprint} \eprint{https://doi.org/10.1137/21M1401243}) \urlprefix\url{https://doi.org/10.1137/21M1401243}

\bibitem{liu2023NonlinearReactionDiffusion}
Liu J~P, An D, Fang D and et~al 2023 {\em Communications in Mathematical Physics\/} {\bf 404} 963--1020 \urlprefix\url{https://doi.org/10.1007/s00220-023-04857-9}

\bibitem{krovi2023ImproveLinearNonlinearODES}
Krovi H 2023 {\em Quantum\/} {\bf 7} 913

\bibitem{An2023FractionalRxnDiffusion}
An D and Trivisa K 2023 {\em arXiv preprint arXiv:2310.18900\/}

\bibitem{Conjugate_Gradient}
Shewchuk J~R 1994 An introduction to the conjugate gradient method without the agonizing pain \urlprefix\url{https://api.semanticscholar.org/CorpusID:6491967}

\bibitem{Fang2023timemarchingbased}
Fang D, Lin L and Tong Y 2023 {\em {Quantum}\/} {\bf 7} 955 ISSN 2521-327X \urlprefix\url{https://doi.org/10.22331/q-2023-03-20-955}

\bibitem{TrefethenExpOperator}
Schmelzer T and Trefethen L~N 2007 {\em Electronic Transactions on Numerical Analysis\/} {\bf 29} 1+ accessed 5 Dec. 2023 \urlprefix\url{https://link.gale.com/apps/doc/A197802373/AONE?u=anon~c8b15afd&sid=googleScholar&xid=c9b6fb49}

\bibitem{Child_LCU}
Childs A~M and Wiebe N 2012 {\em Quantum Info. Comput.\/} {\bf 12} 901–924 ISSN 1533-7146

\bibitem{Low2018HamiltonianSim_LCU}
Low G~H and Wiebe N 2018 {\em arXiv: Quantum Physics\/} \urlprefix\url{https://api.semanticscholar.org/CorpusID:119402530}

\bibitem{davis1959numerical}
Davis P 1959 On the numerical integration of periodic analytic functions, on numerical approximation, re langer, ed

\bibitem{davis2014methods}
Davis P, Rabinowitz P and Rheinbolt W 2014 {\em Methods of Numerical Integration\/} Computer Science and Applied Mathematics (Elsevier Science) ISBN 9781483264288 \urlprefix\url{https://books.google.com/books?id=mbLiBQAAQBAJ}

\bibitem{IBM_UtilityPaper}
Kim Y, Eddins A, Anand S, Wei K~X, van~den Berg E, Rosenblatt S, Nayfeh H, Wu Y, Zaletel M, Temme K and Kandala A 2023 {\em Nature\/} {\bf 618} 500--505 \urlprefix\url{https://doi.org/10.1038/s41586-023-06096-3}

\bibitem{Google_Supremacy}
Arute F {\em et~al.\/} 2019 {\em Nature\/} {\bf 574} 505--510 \urlprefix\url{https://doi.org/10.1038/s41586-019-1666-5}

\bibitem{cerezo2021VQA}
Cerezo M, Arrasmith A, Babbush R, Benjamin S~C, Endo S, Fujii K, McClean J~R, Mitarai K, Yuan X, Cincio L {\em et~al.\/} 2021 {\em Nature Reviews Physics\/} {\bf 3} 625--644

\bibitem{epstein1974variation}
Epstein S 1974 {\em The Variation Method in Quantum Chemistry\/} 1st ed (Academic Press) ISBN 9780323157476

\bibitem{OriginalVQE}
Peruzzo A, McClean J, Shadbolt P, Yung M~H, Zhou X~Q, Love P, Aspuru-Guzik A and O’Brien J~L 2019 {\em Nature Communications\/} \urlprefix\url{https://doi.org/10.1038/ncomms5213}

\bibitem{VQE_Review}
Tilly J, Chen H, Cao S, Picozzi D, Setia K, Li Y, Grant E, Wossnig L, Rungger I, Booth G~H and Tennyson J 2022 {\em Physics Reports\/} {\bf 986} 1–128 ISSN 0370-1573 \urlprefix\url{http://dx.doi.org/10.1016/j.physrep.2022.08.003}

\bibitem{cao2019_qChemqComp}
Cao Y, Romero J, Olson J~P, Degroote M, Johnson P~D, Kieferová M, Kivlichan I~D, Menke T, Peropadre B, Sawaya N~P~D, Sim S, Veis L and Aspuru-Guzik A 2019 {\em Chemical Reviews\/} {\bf 119} 10856--10915 \urlprefix\url{https://doi.org/10.1021/acs.chemrev.8b00803}

\bibitem{kandala2017_IBM_VQE}
Kandala A, Mezzacapo A, Temme K, Takita M, Brink M, Chow J~M and Gambetta J~M 2017 {\em Nature\/} {\bf 549} 242--246 \urlprefix\url{https://www.nature.com/articles/nature23879}

\bibitem{Carlos_VQLS}
Bravo-Prieto C, LaRose R, Cerezo M, Subaşı Y, Cincio L and Coles P~J 2019 {\em Quantum\/} \urlprefix\url{https://api.semanticscholar.org/CorpusID:219924287}

\bibitem{Huang2021NearTermQuantumAlgorithms}
Huang H~Y, Bharti K and Rebentrost P 2021 {\em New Journal of Physics\/} {\bf 23}

\bibitem{Anschuetz2022_VQEChallenges}
Anschuetz E~R and Kiani B~T 2022 {\em Nature Communications\/} {\bf 13}

\bibitem{yuan2019_varQITE_theory}
Yuan X, Endo S, Zhao Q, Li Y and Benjamin S~C 2019 {\em Quantum\/} {\bf 3} 191

\bibitem{mcardle2019variational}
McArdle S, Jones T, Endo S, Li Y, Benjamin S~C and Yuan X 2019 {\em npj Quantum Information\/} {\bf 5} 75

\bibitem{MottaQITE}
Motta M, Sun C, Tan A~T, O’Rourke M~J, Ye E, Minnich A~J, Brand{\~a}o F~G~S~L and Chan G~K~L 2019 {\em Nature Physics\/} {\bf 16} 205--210 \urlprefix\url{https://api.semanticscholar.org/CorpusID:208174822}

\bibitem{Ball2006_MaxwellEq_Theory}
Ball D~W 2006 Field guide to spectroscopy {\em SPIE PRESS BOOK\/} (The International Society for Optics and Photonics) \urlprefix\url{https://api.semanticscholar.org/CorpusID:118576773}

\bibitem{Gaitonde2006MagnetohydrodynamicEP}
Gaitonde D~V 2006 {\em Journal of Propulsion and Power\/} {\bf 22} 498--510 \urlprefix\url{https://api.semanticscholar.org/CorpusID:123057782}

\bibitem{gaitonde2008highspeed}
Gaitonde D~V 2008 {\em J. Propuls. Power\/} {\bf 24} 946--961

\bibitem{michael2019multiphysics}
Michael L, Millmore S~T and Nikiforakis N 2019 {\em Commun. Appl. Math. Comput.\/}

\bibitem{Shang2001RadioBlackout}
Shang J~J~S 2001 {\em Progress in Aerospace Sciences\/} {\bf 37} 1--20 \urlprefix\url{https://api.semanticscholar.org/CorpusID:108835178}

\bibitem{MHD_Propulsion}
Burton R~L and Turchi P 1998 {\em Journal of Propulsion and Power\/} {\bf 14} 716--735

\bibitem{otin2019computational}
Otin R, Aria S, Thompson V, Lobel R, Williams J, Vizvary Z, Iglesias D and Porton M 2019 {\em NAFEMS World Congress 2019\/} Submitted, Quebec City, Canada, 17-20 June 2019

\bibitem{Wolski2011TheoryOE}
Wolski A 2011 {\em arXiv: Accelerator Physics\/} \urlprefix\url{https://api.semanticscholar.org/CorpusID:119219860}

\bibitem{Brackbill1980_DivConstraint}
Brackbill J~U and Barnes D~C 1980 {\em Journal of Computational Physics\/} {\bf 35} 426--430 \urlprefix\url{https://api.semanticscholar.org/CorpusID:119880253}

\bibitem{MikeIke}
Nielsen M~A and Chuang I~L 2011 {\em Quantum Computation and Quantum Information: 10th Anniversary Edition\/} 10th ed (USA: Cambridge University Press) ISBN 1107002176

\bibitem{Alghassi2022VarQITE}
Alghassi H, Deshmukh A, Ibrahim N, Robles N, Woerner S and Zoufal C 2022 {\em {Quantum}\/} {\bf 6} 730 ISSN 2521-327X \urlprefix\url{https://doi.org/10.22331/q-2022-06-07-730}

\bibitem{VQA_Optimizers_Benchmark}
Bonet-Monroig X, Wang H, Vermetten D, Senjean B, Moussa C, Back T, Dunjko V and O’Brien T~E 2021 {\em Physical Review A\/} \urlprefix\url{https://api.semanticscholar.org/CorpusID:244709425}

\bibitem{McClean_2018_Barren_plateaus}
McClean J~R, Boixo S, Smelyanskiy V~N, Babbush R and Neven H 2018 {\em Nature Communications\/} {\bf 9} ISSN 2041-1723 \urlprefix\url{http://dx.doi.org/10.1038/s41467-018-07090-4}

\bibitem{StatsAnalysis}
Ross S~M 2023 8 - statistical analysis of simulated data {\em Simulation (Sixth Edition)\/} ed Ross S~M (Boston: Academic Press) pp 133--149 sixth edition ed ISBN 978-0-323-85739-0 \urlprefix\url{https://www.sciencedirect.com/science/article/pii/B9780323857383000134}

\bibitem{VQE_Measurements_DaveWecker}
Wecker D, Hastings M~B and Troyer M 2015 {\em Phys. Rev. A\/} {\bf 92}(4) 042303 \urlprefix\url{https://link.aps.org/doi/10.1103/PhysRevA.92.042303}

\bibitem{hoffmannV1}
Hoffmann K and Chiang S 2000 {\em Computational Fluid Dynamics\/} ({\em Computational Fluid Dynamics\/} no v. 1) (Engineering Education System) ISBN 9780962373107 \urlprefix\url{https://books.google.com/books?id=98gjAAAACAAJ}

\bibitem{hoffmannV2}
Hoffmann K and Chiang S 2000 {\em Computational Fluid Dynamics\/} ({\em Computational Fluid Dynamics\/} no v. 2) (Engineering Education System) ISBN 9780962373138 \urlprefix\url{https://books.google.com/books?id=-cgjAAAACAAJ}

\bibitem{hoffmannV3}
Hoffmann K and Chiang S 2000 {\em Computational Fluid Dynamics\/} ({\em Computational Fluid Dynamics\/} no v. 3) (Engineering Education System) ISBN 9780962373169 \urlprefix\url{https://books.google.com/books?id=-MgjAAAACAAJ}

\bibitem{LeVequeFDM}
LeVeque R~J 2007 {\em Finite Difference Methods for Ordinary and Partial Differential Equations\/} (Society for Industrial and Applied Mathematics) (\textit{Preprint} \eprint{https://epubs.siam.org/doi/pdf/10.1137/1.9780898717839}) \urlprefix\url{https://epubs.siam.org/doi/abs/10.1137/1.9780898717839}

\bibitem{ThompsonIEEE}
Thompson R~J and Moeller T 2023 {\em IEEE Transactions on Plasma Science\/} {\bf 51} 613--620

\bibitem{holmes2022Expressibility}
Holmes Z, Sharma K, Cerezo M and Coles P~J 2022 {\em PRX Quantum\/} {\bf 3} 010313

\bibitem{Li_2021_prep_periodic_func}
Li J and Kais S 2021 {\em New Journal of Physics\/} {\bf 23} 103022 \urlprefix\url{https://dx.doi.org/10.1088/1367-2630/ac2cb4}

\bibitem{Rattew2021efficient}
Rattew A~G, Sun Y, Minssen P and Pistoia M 2021 {\em {Quantum}\/} {\bf 5} 609 ISSN 2521-327X \urlprefix\url{https://doi.org/10.22331/q-2021-12-23-609}

\bibitem{2020SciPy-NMeth}
Virtanen P, Gommers R, Oliphant T~E, Haberland M, Reddy T, Cournapeau D, Burovski E, Peterson P, Weckesser W, Bright J, {van der Walt} S~J, Brett M, Wilson J, Millman K~J, Mayorov N, Nelson A~R~J, Jones E, Kern R, Larson E, Carey C~J, Polat {\.I}, Feng Y, Moore E~W, {VanderPlas} J, Laxalde D, Perktold J, Cimrman R, Henriksen I, Quintero E~A, Harris C~R, Archibald A~M, Ribeiro A~H, Pedregosa F, {van Mulbregt} P and {SciPy 10 Contributors} 2020 {\em Nature Methods\/} {\bf 17} 261--272

\bibitem{QiskitCommunity2017}
{Qiskit Community} 2017 Qiskit: {{An}} open-source framework for quantum computing \urlprefix\url{https://github.com/Qiskit/qiskit}

\end{thebibliography}

\end{document}